\documentclass[twocolumn]{aastex63}
\usepackage{graphicx}
\usepackage{natbib}
\usepackage{url}
\usepackage{footnote}
\usepackage{amsmath,amssymb}
\usepackage{cleveref}
\usepackage{float} %required for the placement specifier H
\usepackage[caption=false]{subfig}
\received{}
\revised{}
\accepted{}
\shorttitle{ELGs in LAMOST}
\begin{document}
\title{Strong [OIII]$\lambda5007$ emission line compact galaxies in LAMOST DR9: Blueberries, Green Peas and Purple Grapes}
\correspondingauthor{A-Li Luo}
\email{* lal@nao.cas.cn }
\author[0000-0002-5345-4175]{Siqi Liu}
\affiliation{CAS Key Laboratory of Optical Astronomy, National Astronomical Observatories, Beijing 100101, China}
\affiliation{University of Chinese Academy of Sciences, Beijing 100049, China}
\author[0000-0001-7865-2648]{A-Li Luo$^{*}$}
\affiliation{CAS Key Laboratory of Optical Astronomy, National Astronomical Observatories, Beijing 100101, China}
\affiliation{University of Chinese Academy of Sciences, Beijing 100049, China}
\affiliation{College of computer and Information Management \& Institute for Astronomical Science, Dezhou University, Dezhou 253023, China}
\author[0000-0003-2260-7420]{Huan Yang}
\affiliation{CAS Key Laboratory for Researches in Galaxies and Cosmology, University of Science and Technology of China, Chinese Academy of Sciences, Hefei, Anhui 230026, China}
\affiliation{Las Campanas Observatory, Carnegie Institution of Washington, Casilla 601, La Serena, Chile}
\author[0000-0002-3073-5871]{Shi-Yin Shen}
\affiliation{Key Laboratory for Research in Galaxies and Cosmology, Shanghai Astronomical Observatory, Chinese Academy of Sciences, 80 Nandan Road, Shanghai,  200030, China}
\affiliation{Key Lab for Astrophysics, Shanghai, 200034, China}
\author[0000-0002-4419-6434]{Jun-Xian Wang}
\affiliation{CAS Key Laboratory for Researches in Galaxies and Cosmology, University of Science and Technology of China, Chinese Academy of Sciences, Hefei, Anhui 230026, China}
\affiliation{School of Astronomy and Space Science, University of Science and Technology of China, Hefei 230026, China}
\author[0000-0002-6617-5300]{Hao-Tong Zhang}
\affiliation{CAS Key Laboratory of Optical Astronomy, National Astronomical Observatories, Beijing 100101, China}
\author[0000-0002-9634-2923]{Zhenya Zheng}
\affiliation{Key Laboratory for Research in Galaxies and Cosmology, Shanghai Astronomical Observatory, Chinese Academy of Sciences, 80 Nandan Road, Shanghai, 200030, China}
\affiliation{Key Lab for Astrophysics, Shanghai, 200034, China}
\author[0000-0001-7255-5003]{Yi-Han Song}
\affiliation{CAS Key Laboratory of Optical Astronomy, National Astronomical Observatories, Beijing 100101, China}
\author[0000-0001-8011-8401]{Xiao Kong}
\affiliation{CAS Key Laboratory of Optical Astronomy, National Astronomical Observatories, Beijing 100101, China}

\author[0000-0001-8274-3980]{Jian-Ling Wang}
\affiliation{CAS Key Laboratory of Optical Astronomy, National Astronomical Observatories, Beijing 100101, China}
\author{Jian-Jun Chen}
\affiliation{CAS Key Laboratory of Optical Astronomy, National Astronomical Observatories, Beijing 100101, China}
\affiliation{University of Chinese Academy of Sciences, Beijing 100049, China}
%% Note that the \and command from previous versions of AASTeX is now
%% depreciated in this version as it is no longer necessary. AASTeX
%% automatically takes care of all commas and "and"s between authors names.

%% AASTeX 6.3 has the new \collaboration and \nocollaboration commands to
%% provide the collaboration status of a group of authors. These commands
%% can be used either before or after the list of corresponding authors. The
%% argument for \collaboration is the collaboration identifier. Authors are
%% encouraged to surround collaboration identifiers with ()s. The
%% \nocollaboration command takes no argument and exists to indicate that
%% the nearby authors are not part of surrounding collaborations.

%% Mark off the abstract in the ``abstract'' environment.
\begin{abstract}
Green Pea and Blueberry galaxies are well-known for their compact size, low mass, strong emission lines and analogs to high-$z$ Ly$\alpha$ emitting galaxies.
In this study, 1547 strong [OIII]$\lambda5007$ emission line compact galaxies with 1694 spectra are selected from LAMOST DR9 at the redshift range from 0.0 to 0.59. 
According to the redshift distribution, these samples can be separated into three groups: Blueberries, Green Peas and Purple Grapes. 
Optical [MgII]$\lambda2800$ line feature, BPT diagram, multi-wavelength SED fitting, MIR color, and MIR variability are deployed to identify 23 AGN candidates from these samples, which are excluded for the following SFR discussions.
We perform the multi-wavelength SED fitting with GALEX UV and WISE MIR data.
Color excess from Balmer decrement shows these strong [OIII]$\lambda5007$ emission line compact galaxies are not highly reddened.  
The stellar mass of the galaxies is obtained by fitting LAMOST calibrated spectra with the emission lines masked.  We find that the SFR is increasing with the increase of redshift, while for the sources within the same redshift bin, the SFR increases with mass with a similar slope as the SFMS.
These samples have a median metallicity of 12+log(O/H) of 8.10. 
  The metallicity increases with mass, and all the sources are below the mass-metallicity relation.  
The direct-derived $T_e$-based metallicity from the [OIII]$\lambda$4363 line agrees with the empirical N2-based empirical gas-phase metallicity.
Moreover, these compact strong [OIII]$\lambda5007$ are mostly in a less dense environment.
%Further spectroscopic survey and multi-wavelength photometry will help to reveal more properties for the strong [OIII]$\lambda5007$ emission line compact galaxies.
\end{abstract}

%% Keywords should appear after the \end{abstract} command.
%% See the online documentation for the full list of available subject
%% keywords and the rules for their use.
% UAT: Compact dwarf galaxies (281); Emission line galaxies (459); Catalogs (205);
%% From the front matter, we move on to the body of the paper.
%% Sections are demarcated by \section and \subsection, respectively.
%% Observe the use of the LaTeX \label
%% command after the \subsection to give a symbolic KEY to the
%% subsection for cross-referencing in a \ref command.
%% You can use LaTeX's \ref and \label commands to keep track of
%% cross-references to sections, equations, tables, and figures.
%% That way, if you change the order of any elements, LaTeX will
%% automatically renumber them.
%%
%% We recommend that authors also use the natbib \citep
%% and \citet commands to identify citations.  The citations are
%% tied to the reference list via symbolic KEYs. The KEY corresponds
%% to the KEY in the \bibitem in the reference list below.
% \section{Introduction}
% \label{sec:intro}
\section{Introduction}
\label{sec:intro}
\begin{figure}[!ht]
  \centering
  \includegraphics[width=\hsize]{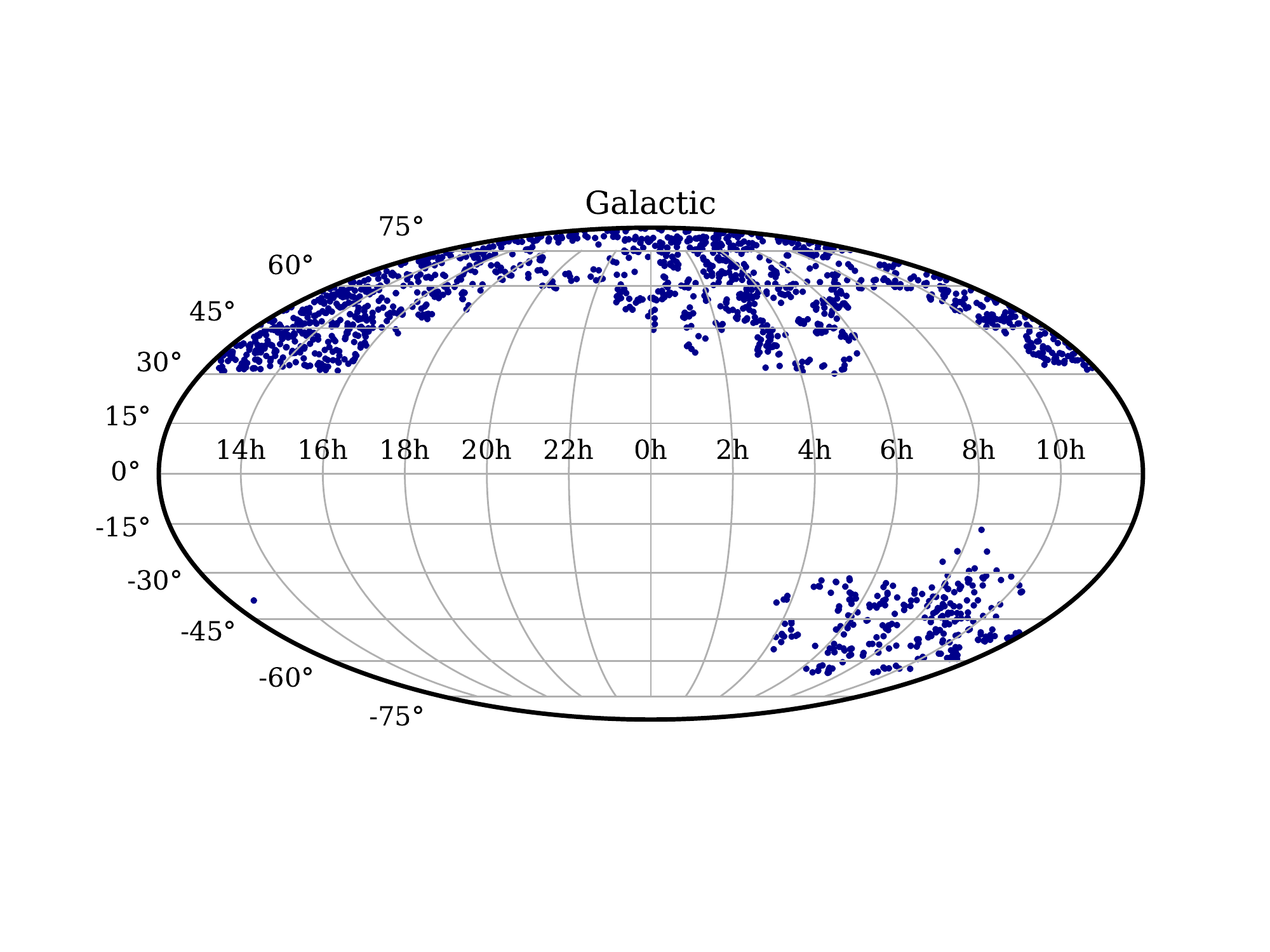}
      \caption{The sources in the Galactic coordinates, mainly around the North Galactic Cap (NGC) and the South Galactic Cap (SGC). }
         \label{fig:galactic-coord}
\end{figure}

The Large sky Area Multi-Object Fiber Spectroscopic Telescope (LAMOST) \citep{1996ApOpt..35.5155W,2004ChJAA...4....1S}, located at the Xinglong Observatory in Hebei Province northeast of Beijing, has an effective aperture of 4 m.  
Ensembling 4000 fibers, the LAMOST facility can collect multiple spectra at the same time\citep{2012RAA....12.1197C}.
% There are two major surveys being conducted at LAMOST: LEGUE (LAMOST Experiment for Galactic Understanding and Exploration survey) and LEGAS (LAMOST ExtraGAlactic Survey). 
The science goal for LAMOST extragalactic survey is to investigate the extragalactic objects for galaxies and QSOs \citep{2012RAA....12..723Z}.
% For the galaxies, the LEGAS survey mainly covers two regions: the North Galactic Cap (NGC) sharing the same footprint as SDSS aims to observe the missing spectra of SDSS due to fiber collisions; and the South Galactic Cap (SGC) aiming to observe a complete sample of galaxies up to $r=18$, therefore the LEGAS survey in the SGC only focus on a strip ($45^{\circ} < \alpha < 60^{\circ}$ and $0.5^{\circ} < \delta < 9.5^{\circ}$)\citep{2015RAA....15.1095L}. 
Fig.\ref{fig:galactic-coord} demonstrates the our selected strong [OIII]$\lambda5007$ emission line compact galaxies in the Galactic coordinates, which will be illustrated in Section 2.2 in detail.  

Green Pea galaxies were first identified in the Galaxy Zoo project \citep{2009MNRAS.399.1191C} by their unique compact size and green color caused by strong [OIII]$\lambda$5007 emission lines. \citet{2009MNRAS.399.1191C} found 251 Green Pea galaxies from SDSS DR 7 by photometric color criteria, where 80 of the sources have spectroscopic data. Following studies have enlarged the sample size for the Green Pea galaxies.\citet{2011ApJ...728..161I} contains 803 star-forming luminous compact galaxies (LCGs) with high H$\beta$ luminosities in the redshift range $z=0.02-0.63$ from SDSS DR7 spectroscopic data.  \citet{2019ApJ...872..145J} contains 800+ Green Pea galaxies from the spectroscopic database of SDSS DR13 and use these samples to to do direct $T_e$-based metallicity calibration.

Blueberry galaxies at $z<0.05$ are the lowest mass young starburst galaxies \citep{2017ApJ...847...38Y} where the [OIII]$\lambda$5007 emission line is located within the $g-$band and the H$\alpha$ emission line is located within the $r-$band.
Purple Grape galaxies\citep{2011ApJ...728..161I,2020ApJ...898...68B} where either that the [OIII]$\lambda$5007 emission line is located in the $i-$band at $z>0.36$ and the UV continuum is redshifted to the $g-$band, or the [OIII]$\lambda$5007 is located in the $g$-band and the H$\alpha$ emission line is located in the $i-$band ($0.112<z<0.36$).
% These galaxies are of general interest because of its ongoing star formation and 

In this work, we compile a large catalog consisting of strong [OIII]$\lambda5007$ emission line compact galaxies with 1694 spectra from LAMOST DR9 at the redshift range from 0.0 to 0.722.
Among the sources 219 have SDSS spectra detections from SDSS DR16 \citep{2017AJ....154...28B,2020ApJS..249....3A}.
Joint with multi-band photometry, we can systematically learn about the properties for these high SFR, compact galaxies spanning a wide range of redshift coverage.  

In Section \ref{sec:sample}, we describe sample selection criteria, how we do the flux calibration of the LAMOST spectra and the mulltiwavelength dataset we use for following discussions.
In Section \ref{sec:AGN}, we describe multi-wavelength SED fitting result, and how we identify the AGN candidates from the samples.
In Section \ref{sec:physical_prop}, we discuss the physical properties for these sources, including the stellar mass of the galaxies from spectral fitting, the SFR, the gas-phase metallicity, and the environment. 
In Section \ref{sec:result}, we summarize the results addressed in this work.

In this work, we use \textit{Wilkinson Microwave Anisotropy Probe} (WMAP) 9 cosmology \citep{2013ApJS..208...19H}, AB magnitudes \citep{1983ApJ...266..713O},
Chabrier IMF \citep{2003PASP..115..763C}, and \citet{bruzual2003stellar} SPS models.

\section{Sample and data}
\label{sec:sample}

\subsection{Sample}
There are mainly two ranges that we select the strong [OIII]$\lambda5007$ samples from the LAMOST DR9 of the extragalactic survey based on the color selection: from both dedicated or non-dedicated ``Green Pea" targets of the input-catalog.

The dedicated ``Green Pea" targets in the input catalog originate from the PI project of LAMOST extragalactic survey add-on program covering a large area of the North Galactic Cap and a strip in the South Galactic Cap.  %They are selected by color criteria in SDSS DR12. 
%Bright sources are assigned with high priority.  
By now, 2309 spectra have been observed in LAMOST DR9\footnote{\url{http://www.lamost.org/dr9/}. 
% The initial input catalog of candidate strong [OIII]$\lambda5007$ emission-line galaxies are determined from SDSS $ugriz$ colors from SDSS DR12\footnote{\url{https://www.sdss.org/dr12/}} CasJobs system\footnote{\url{http://skyserver.sdss.org/casjobs/}}.
The initial color criteria determined from SDSS 12\footnote{\url{https://www.sdss.org/dr12/}} $ugriz$ photometry are as follows.
% The cModelMag\_r of these candidates are in the range of 16 to 20.5 mag.  
For the sources in $z<0.12$, where the [OIII]$\lambda5007$ is located within the $g-$band, the criteria are two-fold: loose color criteria with $u-g \le 0.3$, $r-g \le 0.1$ and $i-g \le -0.7$ but require the \texttt{type}=3 to be galaxy; or no constraint on morphology but require strict color criteria with $u-g \le 0.5$, $r-g \le 0.5$.
For the sources in $0.12 < z < 0.4$, where the [OIII]$\lambda5007$ is located within the $r-$band, the criteria are displayed in Equation 1 (the same as Equations (1)-(5) in \citet{2009MNRAS.399.1191C}).  
% together with photometric flags to ensure valid detections.
}
\begin{eqnarray}
     u-r&\le& 2.5, \nonumber\\
     r-i &\le& -0.2, \nonumber\\
     r-z &\le& 0.5,\\
     g-r &\ge& r-i+0.5, \nonumber\\
     u-r &\le& 2.5(r-z). \nonumber
        \label{eq:color} 
\end{eqnarray}

% For the candidates with lower redshift ($z<0.1$) and situating their [OIII]$\lambda5007$ lines within $g-$band, we select them using the $u-g-r$ color together with the morphology in a two-fold strategy (that either use strict color criteria without restricting the morphology or by limiting the source to be classified as galaxy from SDSS pipeline and apply loose criteria on color).
% The color criteria for the low$-z$ candidates are as follows \citep{2017ApJ...847...38Y}:
% \begin{equation}
%     \begin{split}
%     \rm g-r & < 0.5 ~\rm and ~r-i  < 1.0, ~\rm and ~g-i  < -0.5.\\
%     \rm g-r &  < -0.7 ~\rm and ~g-i  < -1.0 ~\rm and ~g-u  < -0.3.
%     \end{split}
% \end{equation}
% We display this criteria in the left panel of Fig.\ref{fig:color} with darkblue solid lines.
% The Green Pea selection criteria according to \citet{2009MNRAS.399.1191C} is as follows:

\begin{figure}[!ht]
  \centering
  \includegraphics[width=\hsize]{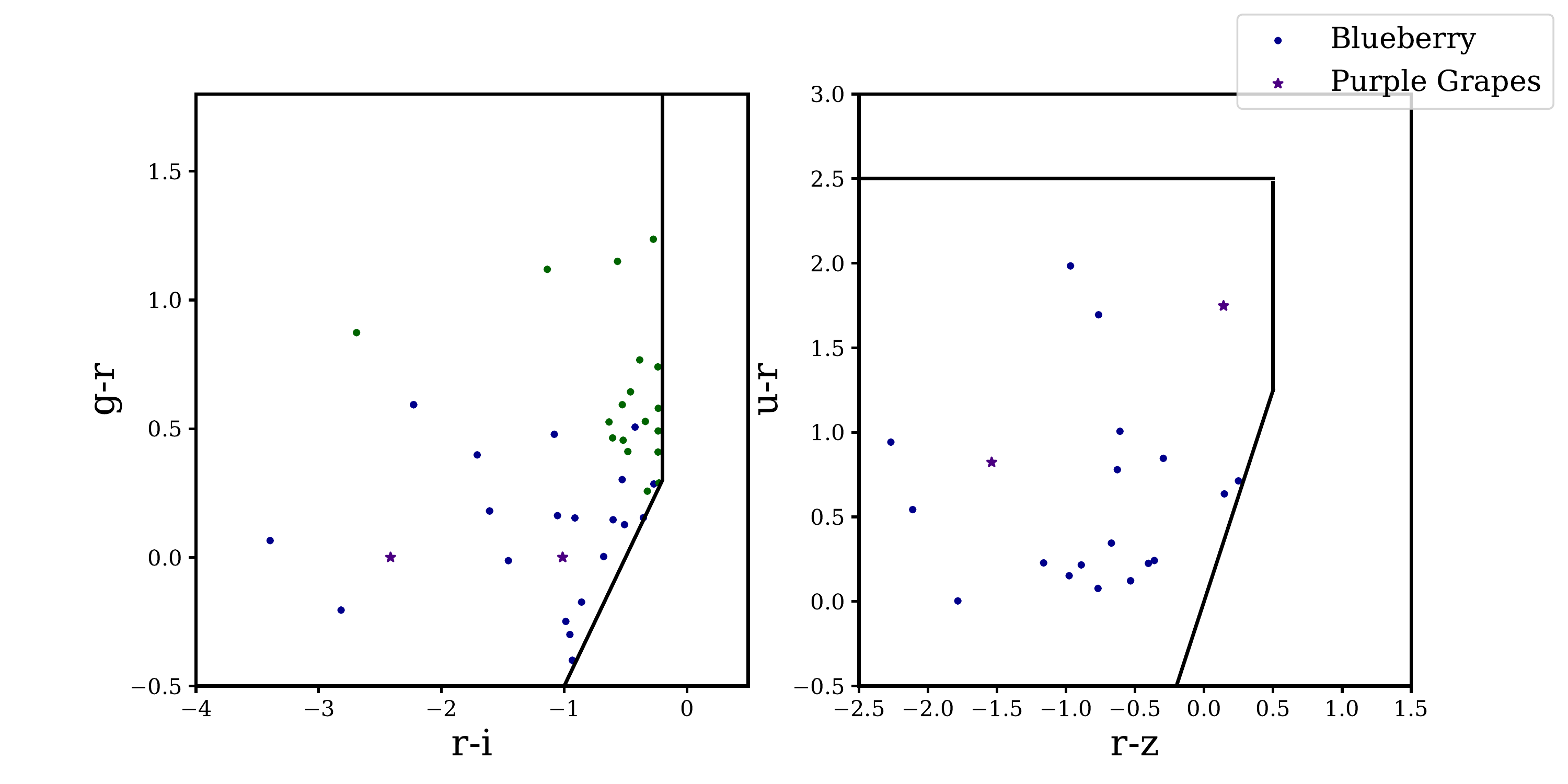}
      \caption{44 newly selected Green Pea/Blueberry/Purple Grape galaxies samples from the color selection criteria.
     The green solid lines mark the color selection criteria from \citet{2009MNRAS.399.1191C} of Green Pea galaxies.
     There are 3 Blueberry galaxies and 1 Green Pea galaxy located at the left part of the left panel plot which are not displayed in this figure, and there are four Blueberry galaxies located at the left part in the right panel not displayed.
     }
         \label{fig:color}
  \end{figure}

For the non-dedicated ``Green Pea" targets in the input-catalog (by non-dedicated we mean that the initial purpose for observing this target was not intended for observing Green Pea galaxies, Blueberry galaxies or other compact emission line galaxies), we select extra 23 Blueberry galaxies (25 spectra), 19 Green Pea galaxies (19 spectra) and 2 Purple Grape galaxies (3 spectra) using the same color criteria in Equation 1, as displayed in Fig.\ref{fig:color}.

Next, we remove the sources with a large radius in $r-$band whose $\texttt{petroRad\_r}>5~\rm{arcsec}$.  
We only keep the sources where the \texttt{mode} keyword is equal to 1 from the PhotoObjAll in SDSS DR13 which means primary object.
Furthermore, we visually inspect the SDSS images and only keep the isolated sources.
For the last step, we visually inspect each spectrum to ensure the strong [OIII]$\lambda5007$ emission lines lie within the spectra and only keep the sources where the flux of [OIII]$\lambda5007$ line is above $\rm 3\times10^{17} erg/s/cm^2$.

Finally,  1547 in total strong [OIII]$\lambda5007$ emission line compact galaxies with 1694 spectra are selected.
LAMOST pipeline have classified these sources into four types: 1646 \texttt{GALAXY}, 3 \texttt{STAR}, 35 \texttt{QSO} and 14 \texttt{Unknown}.    

\begin{figure}[!ht]
  \centering
  \includegraphics[width=\hsize]{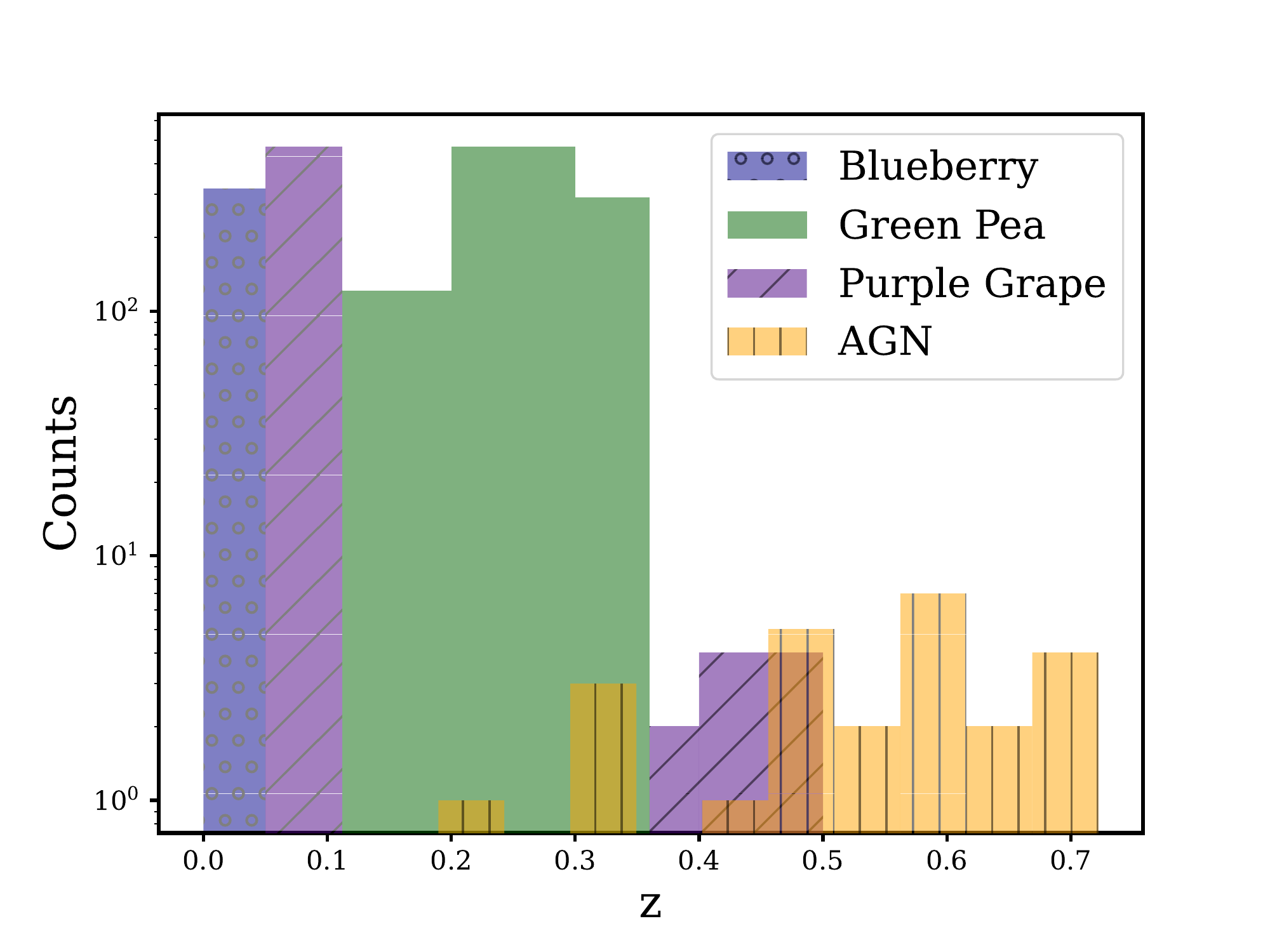}
      \caption{The spectral redshift distribution of the sources.}
         \label{fig:zdist}
\end{figure}

\begin{figure}[!ht]
  \centering
  \includegraphics[width=\hsize]{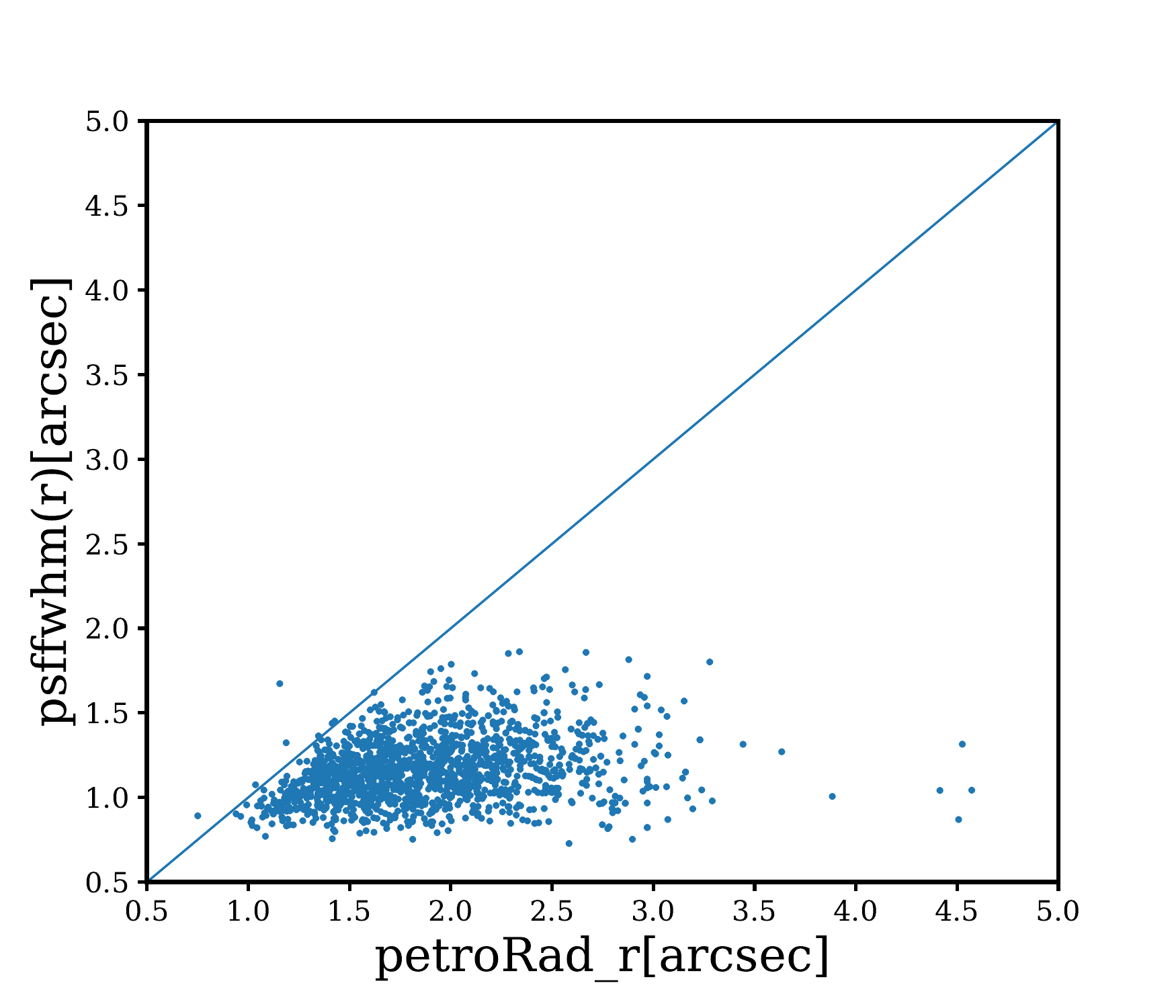}
      \caption{The scatter plot of the $\texttt{petroRad\_r}$ and $\texttt{psfFWHM(r)}$ of our strong [OIII]$\lambda5007$ emission line compact galaxies. }
         \label{fig:rad_psffwhm}
\end{figure}

\begin{figure}[!ht]
  \centering
  \includegraphics[width=\hsize]{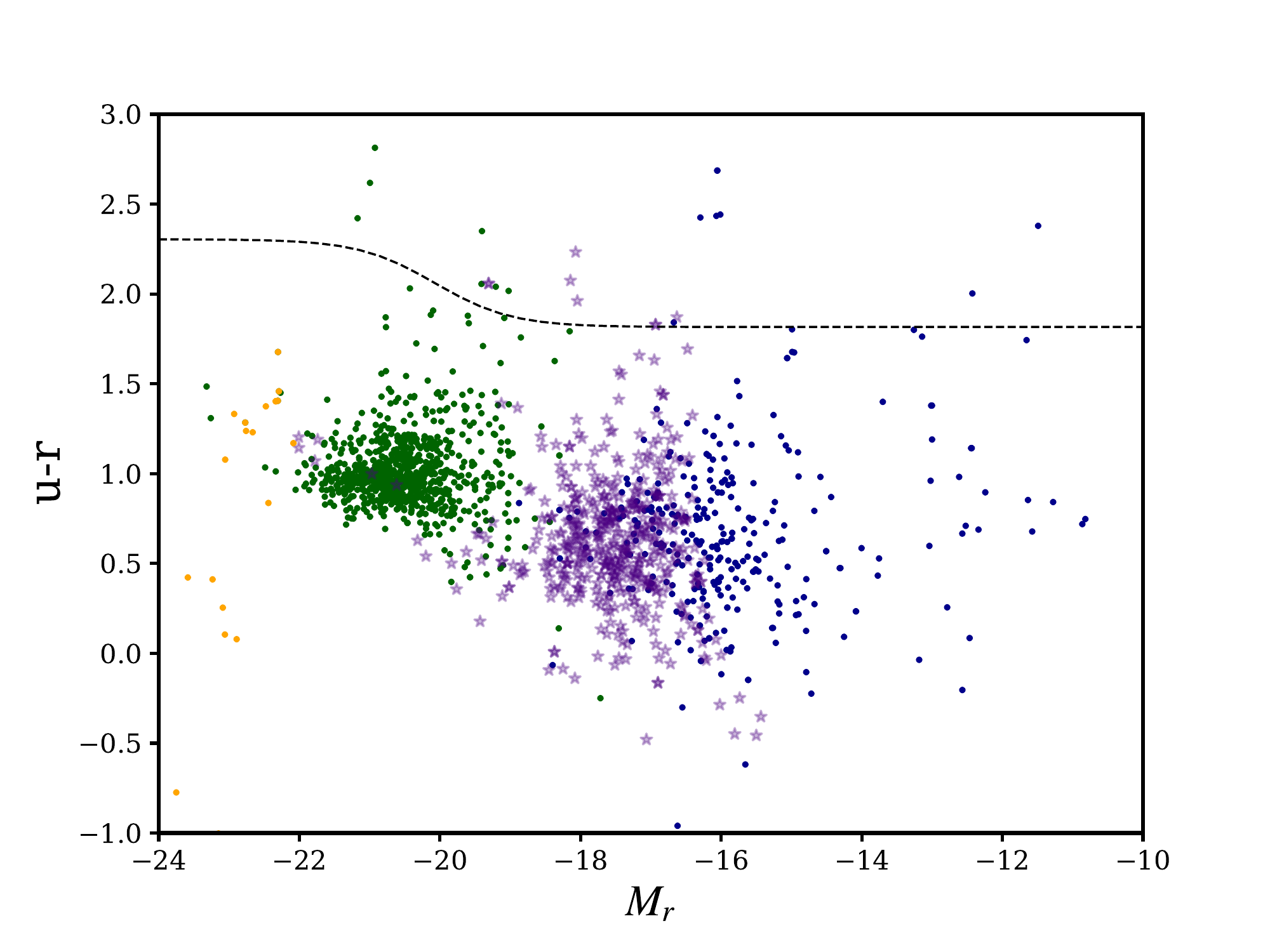}
      \caption{Color-magnitude diagram of the strong [OIII]$\lambda5007$ emission line compact galaxies.  The absolute magnitude of the $r-$band and the $u-r$ color are k-corrected with \citet{2010MNRAS.405.1409C,2012MNRAS.419.1727C}.  The dark blue dots mark the Blueberry galaxies,  the dark green dots mark the Green Pea galaxies, the indigo stars mark the Purple Grape galaxies, and the orange dots mark the spectroscopically-confirmed AGNs.  The majority of our samples locate in the blue could region of the color-magnitude diagram, below the dividing line defined by \citet{2004ApJ...600..681B} between the red sequence and the blue cloud.    }
         \label{fig:CMD}
\end{figure}

Fig.\ref{fig:zdist} shows the spectral redshift distribution of these samples.
The majority of our sources are un-resolved or barely resolved, where the Petrosian radius of the source is close to the FWHM of the PSF as displayed in Fig.\ref{fig:rad_psffwhm}.  We display our sources in the $u-r$ vs $M_r$ color-magnitude diagram in Fig.\ref{fig:CMD}.
As we have expected the majority of the strong [OIII]$\lambda5007$ galaxies are located in the blue cloud region of the color-magnitude diagram under the division line defined in  \citet{2004ApJ...600..681B}.

% We place an upper limit of the distance from the source whose $\texttt{petroRad\_r=1.865}$ times luminosity distance at 37.1 kpc (at $z=0.67$) for our sources.

% There are two redshift regions that exhibit the samples that are called Purple Grapes: $0.05<z<0.112$ and $0.36<z<0.722$.
% In the lower redshift region, the strong [OIII]$\lambda5007$ line is in the g-band, while the strong H$\alpha$ emission line locates in the i-band.
% In the higher redshift region, the strong [OIII]$\lambda5007$ line is in the i-band, together with the strong young star components in the g-band, the color of these objects appear to be purple in the $gri$ three-band false color image.
% We call them Purple Grapes, according to the customs in \citet{2020ApJ...898...68B}.
%We figure out 21 AGN candidates with the existence of [MgII]$\lambda$2800 line in the final samples of 1547 strong [OIII]λ5007  emission line compact galaxies. 
All the sources within the redshift range within $0.59<z<0.73$ are AGN candidates, whose selection criteria will be discussed in Section 3.

% There are 21 optical spectra being identified as AGN spectroscopically within our samples.

\subsection{LAMOST spectra flux calibration}
The 16 LAMOST spectrographs are designed to have a theoretical resolution of $R=1800$, covering the wavelength range from $3600-9000\AA$.  
The blue portion spectrograph covers the wavelength range of $3690-5900 \AA$ and the red portion spectrograph spans the wavelength coverage is $5700-9100\AA$ with $200\AA$ coverage overlap \citep{2012RAA....12.1243L}.

\begin{figure}[!ht]
  \centering
  \includegraphics[width=\hsize]{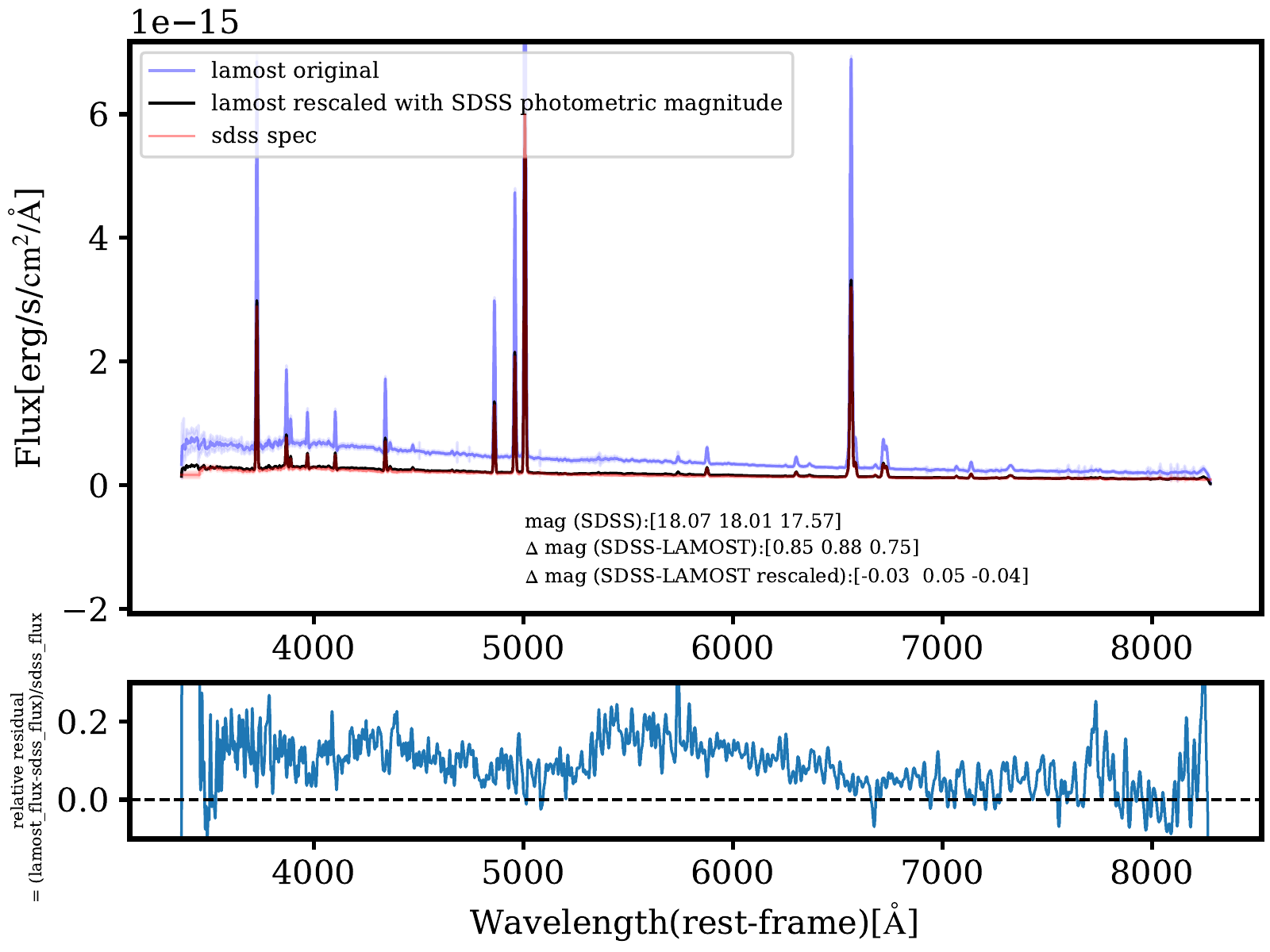}
      \caption{An example of flux calibration of LAMOST spectra.  In the upper panel, we plot the original LAMOST spectra before flux calibration (transparent blue curve), the LAMOST spectra that are re-calibrated with SDSS $gri$ photometric magnitudes (black curve), and the SDSS spectra (red curve).  In the lower panel, we plot the relative residual = (LAMOST\_flux-SDSS\_flux)/SDSS\_flux.  We also mark the 0 level with the horizontal line.  We have smoothed both the LAMOST spectra and the SDSS spectra with a kernel of 3.}
         \label{fig:flux-calibration}
  \end{figure}

The response curves of the 16 spectrographs have been removed from LAMOST spectra, but the spectra flux is not physically calibrated \citep{2016ApJS..227...27D}.
We re-calibrate the LAMOST spectra according to the SDSS $gri$ photometry following \citet{2018MNRAS.474.1873W}  as described below: LAMOST spectra are convolved with the SDSS $gri$ filters \citep{1996AJ....111.1748F} to obtain the synthetic magnitude for these three bands, then compared with the SDSS photometric magnitude. A zeroth-order or first-order polynomial is used to fit the magnitude difference array and apply this correction to the LAMOST spectra. 
An example of the flux calibration result is shown in Fig.\ref{fig:flux-calibration}.
 
For some sources that the LAMOST data reduction pipeline did not calculate their redshift, we need to determine the redshift by fitting the emission lines.

\subsection{UV and MIR photometry}
% \subsubsection{GALEX photometry}
We cross-match the sources with the \textit{GALAXY EVOLUTION EXPLORER} (GALEX) mission \citep{2005ApJ...619L...1M} the revised GALEX catalog of UV sources (GUVcat\_AIS) from GR6+7 (which contains 82,992,086 sources) \citep{2017ApJS..230...24B,Bianchi_2017} using a radius of 3 arcsec and have obtained 1266 cross-matched results.
% We calculate the UV luminosity of the sources with the luminosity distance calculated from the redshift using Astropy package in WMAP9 cosmology. 
% We calculate the median luminosity for these 1266 sources in the FUV is $1.187186\times 10^{15} L_{\odot}$.
% Again this is in agreement with the discussion in \citet{2009MNRAS.399.1191C} that these sources are UV luminous and with low reddening values. 

% \subsubsection{MIR photometry: WISE and Spitzer}
The \textit{Wide-field Infrared Survey Explorer} (WISE) \citep{2010AJ....140.1868W} mission has conducted a MIR survey with 3.4, 4.6, 12, and 22$\micron$ band-passes.
We cross-match our sources in 10 arcsec separation with the AllWISE source catalog, which contains photometry and astrometry of over 747 million objects.
Although this separation is large, we have eye-checked every source to ensure there is no contamination source within the search radius.
We perform the following criteria in SNR and $\chi^2_{\nu}$ space to remove the fake objects from the cross-matched results, as \citet{2014ApJ...791..131K} Section 3.1.1 has performed.  There are 138 sources after cross-matching with WISE.
% \begin{equation}
% \begin{split}
% \rm
%     w1rchi2 &< (w1snr-3)/7, \\
%     w2rchi2 &< 0.1 \times w2snr - 0.3,\\
%     w3rchi2 &< 0.125 \times w3snr - 1,\\
%     w4rchi2 &< 0.2 \times w4snr - 2.
% \end{split}
% \end{equation}
% \begin{eqnarray}
%     \rm w1rchi2 &<&\rm  (w1snr-3)/7, \nonumber\\
%     \rm w2rchi2 &<&\rm  0.1 \times w2snr - 0.3,\\
%     \rm w3rchi2 &<&\rm 0.125 \times w3snr - 1, \nonumber\\
%     \rm w4rchi2 &<&\rm  0.2 \times w4snr - 2. \nonumber
% \end{eqnarray}

% \subsubsection{Spitzer data}
We cross-match with \textit{Spitzer Space Telescope} \citep{2004ApJS..154....1W} data with Spitzer Enhanced Imaging Products (SEIP) using 3 arcsec radius separation and obtained 32 sources that have detections in Infrared Array Camera (IRAC) \citep{2004ApJS..154...10F} and 37 that have detections in Multiband Imaging Photometer for Spitzer (MIPS) \citep{2004ApJS..154...25R}. 
We use Spitzer and WISE photometry for multi-band SED fitting as described in the next subsection.

\section{Multiwaveband  SED fitting and AGN candidates}
\label{sec:AGN}
\subsection{CIGALE SED fitting}
\begin{figure}[!ht]
  \centering
  \includegraphics[width=\hsize]{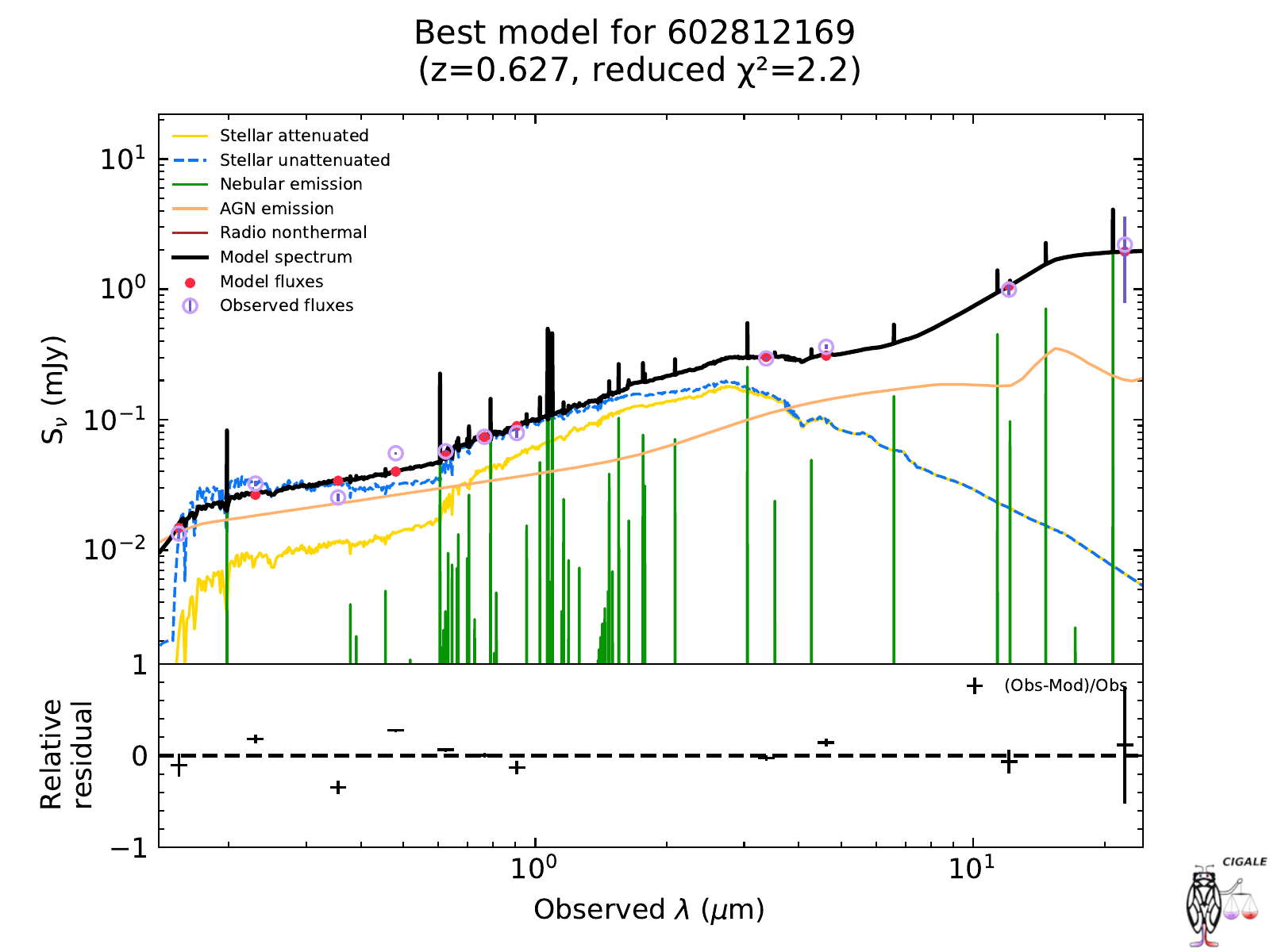}
      \caption{CIGALE SED fitting with GALEX, SDSS $ugriz$ and WISE $W_1$ to $W_4$ photometry.  The dust temperature from the best-fitting result for this source is 325K and the AGN fraction is 20\%.}
         \label{fig:CIGALE-sed}
  \end{figure}

We fit the sources with multi-wavelength photometry with CIGALE \citep{2005MNRAS.360.1413B,2009A&A...507.1793N,2019A&A...622A.103B} trying to use the full photometric data for the 138 sources. As described in the above section, however, GALEX (97 sources) and Spitzer (5 sources) data are available for some of them in addition to SDSS and WISE photometry for the 138 sources.
% Otherwise, we only use SDSS and WISE photometry.
For the configuration of the SED creation modules, we use the delayed$-\tau$ star formation history, BC03 \citep{2003MNRAS.344.1000B}, Chabrier IMF \citep{2003PASP..115..763C}, nebular emission lines, dust attenuated modified starburst model, the dust emission model from \citet{2012MNRAS.425.3094C}, and the \citet{2006MNRAS.366..767F} AGN model.
An example of the CIGALE SED fitting result is shown in Fig.\ref{fig:CIGALE-sed}.

\subsection{AGN candidates}
We identify AGNs from our samples using five aspects: (1) the existence of [MgII]$\lambda2800$ emission line in the optical spectra (as addressed in Section 2.1), (2) the BPT diagram, (3) AGN fraction determination from CIGALE multi-wavelength SED fitting, (4) MIR color, and (5) MIR variability.

\begin{figure}[!ht]
  \centering
  \includegraphics[width=\hsize]{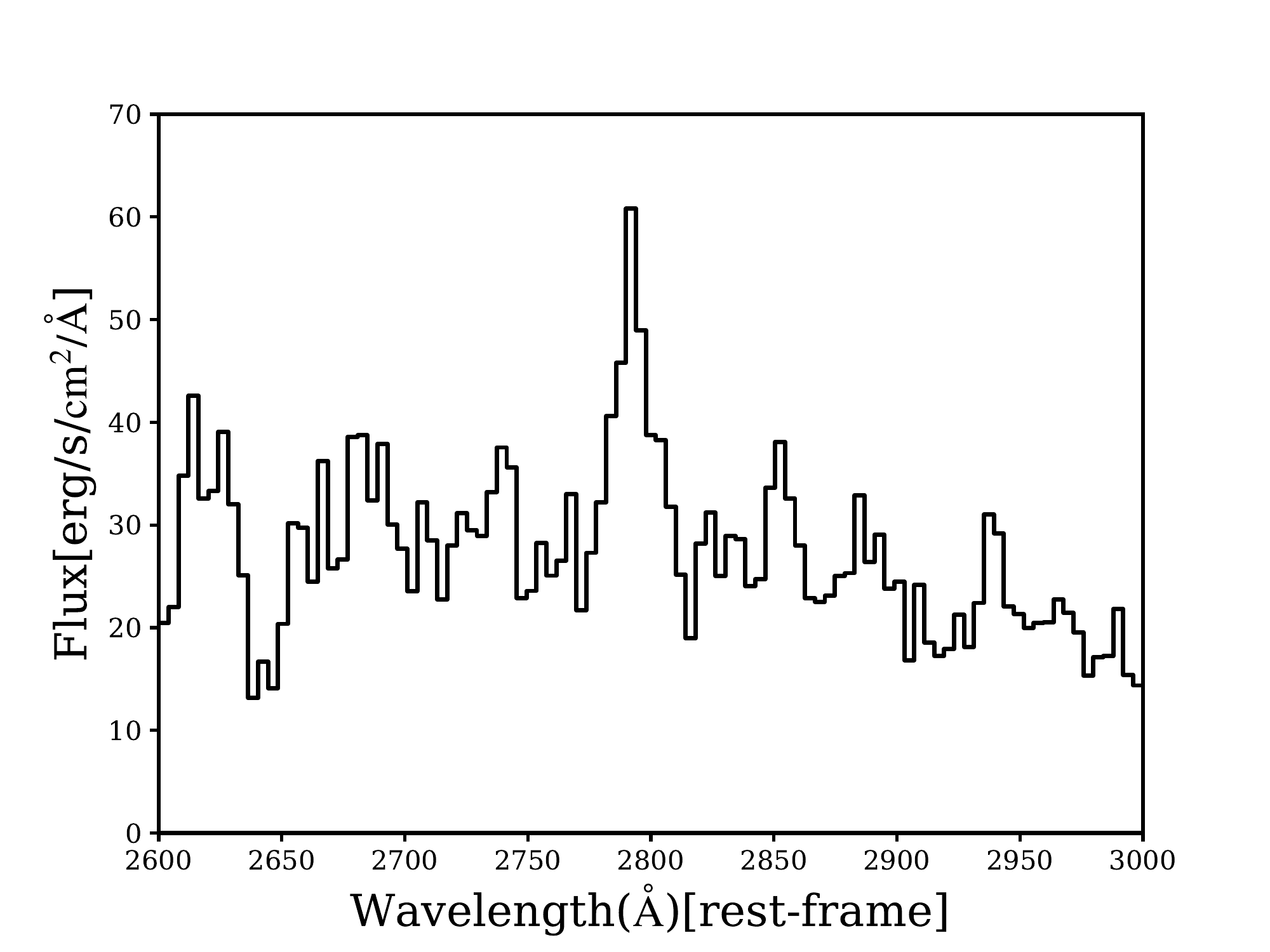}
      \caption{The stacked spectra of [MgII]$\lambda2800$ lines for three epochs with the AGN candidate LAMOST designation J162736.10+56225.8.}
         \label{fig:Mg_3epochs}
\end{figure}

For the optical spectra, we spectroscopically confirm 19 AGNs with 21 spectra from LAMOST DR9 with the existence of [MgII]$\lambda2800$ emission lines.
We display the stacked spectra with three-epoch [MgII]$\lambda$2800 detections in Fig.\ref{fig:Mg_3epochs}, and name the source by its LAMOST designation J162736.10+56225.8, where we have three spectra from LAMOST.
% \begin{figure}[h!]
%   \centering
%   \includegraphics[width=\hsize]{1.2376686809923592e+18three_epochs_OIII_Hb.pdf}
%       \caption{The variability of the [OIII]$\lambda$5007 emission lines, [OIII]$\lambda$5007 emission lines and H$\beta$ emission lines from the LAMOST spectra. 
%       The time separation between the blue (first) and the orange (second) detection is 439 days and the time separation between the orange (second) and the green (third) detection is 1 day.}
%          \label{fig:optical_spec_variable}
% \end{figure}
 
\begin{figure}[!ht]
  \centering
  \includegraphics[width=\hsize]{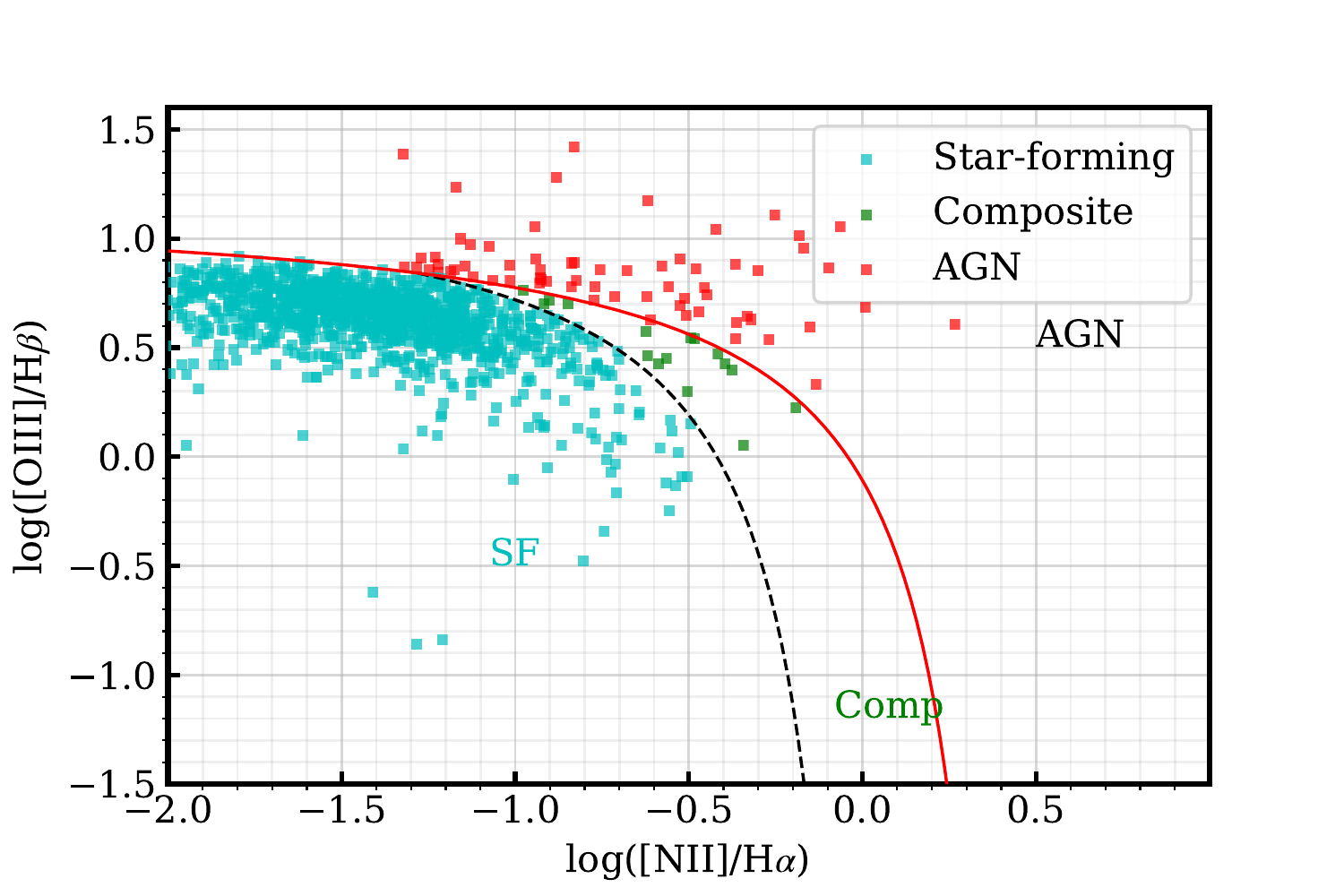}
      \caption{The BPT diagram of the strong [OIII]$\lambda5007$ emission line compact galaxies.  This figure use the plotting routine in \citet{2019AJ....158...74C}.}
         \label{fig:bpt}
  \end{figure}

Using the BPT diagram \citep{1981PASP...93....5B,1987ApJS...63..295V} we classify the galaxies into different spectral types according to the definitions in \citet{2001ApJ...556..121K,2003MNRAS.346.1055K,2006MNRAS.372..961K}.
There are 1423 galaxies (1547 spectra) that are classified as star-forming galaxies, 16 galaxies (16 spectra) as composite, and 71 galaxies (73 spectra) as AGN.  
We plot the distribution for these galaxies in the BPT diagram in Fig.\ref{fig:bpt}.
We do not consider all the sources classified as AGN through the BPT diagram are real AGN sources as the majority of the sources are located in the upper left region of the diagram, where the high ionization is caused by the high [OIII] lines excited from the star-forming region, not the nuclear region.
On the other hand, there might be AGN contaminants located in the SF region especially in the low-metallicity regime as addressed in \citet{2021arXiv210513400H}.

\begin{figure}[!ht]
  \centering
  \includegraphics[width=\hsize]{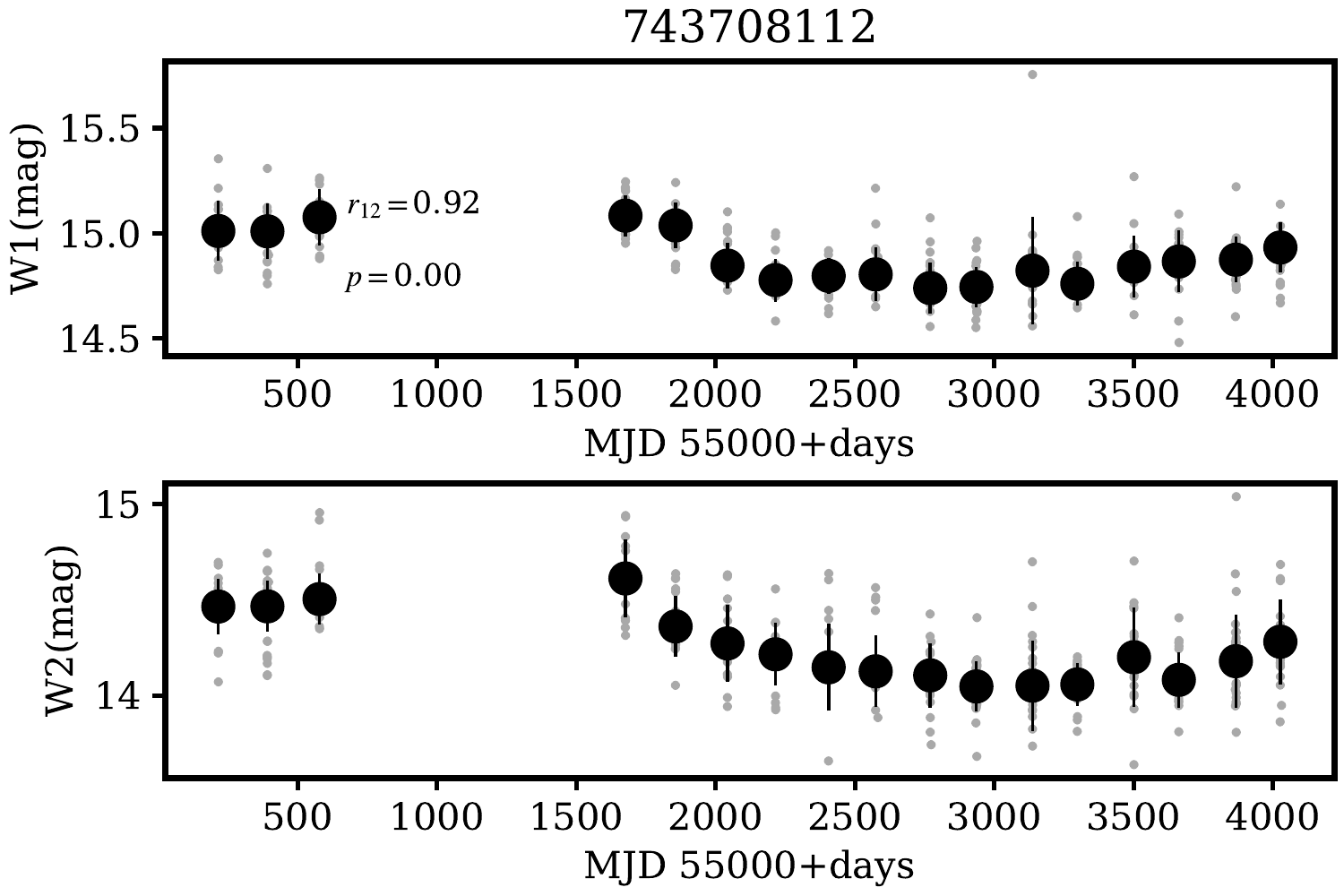}
      \caption{The light curve of one AGN candidate of $W_1$ (upper panel) and $W_2$ (lower panel), where the title marks the obsid from LAMOST.
      The Pearson $r_{12}$ correlation is 0.93 with a p-value of 0.0.
      The single exposure magnitudes are marked with gray scatter points, where the median value and the standard deviation of the exposures from the same epoch within 60 days are marked with black error bars.}
         \label{fig:light_curve_1}
  \end{figure}

\begin{figure}[!ht]
  \centering
  \includegraphics[width=\hsize]{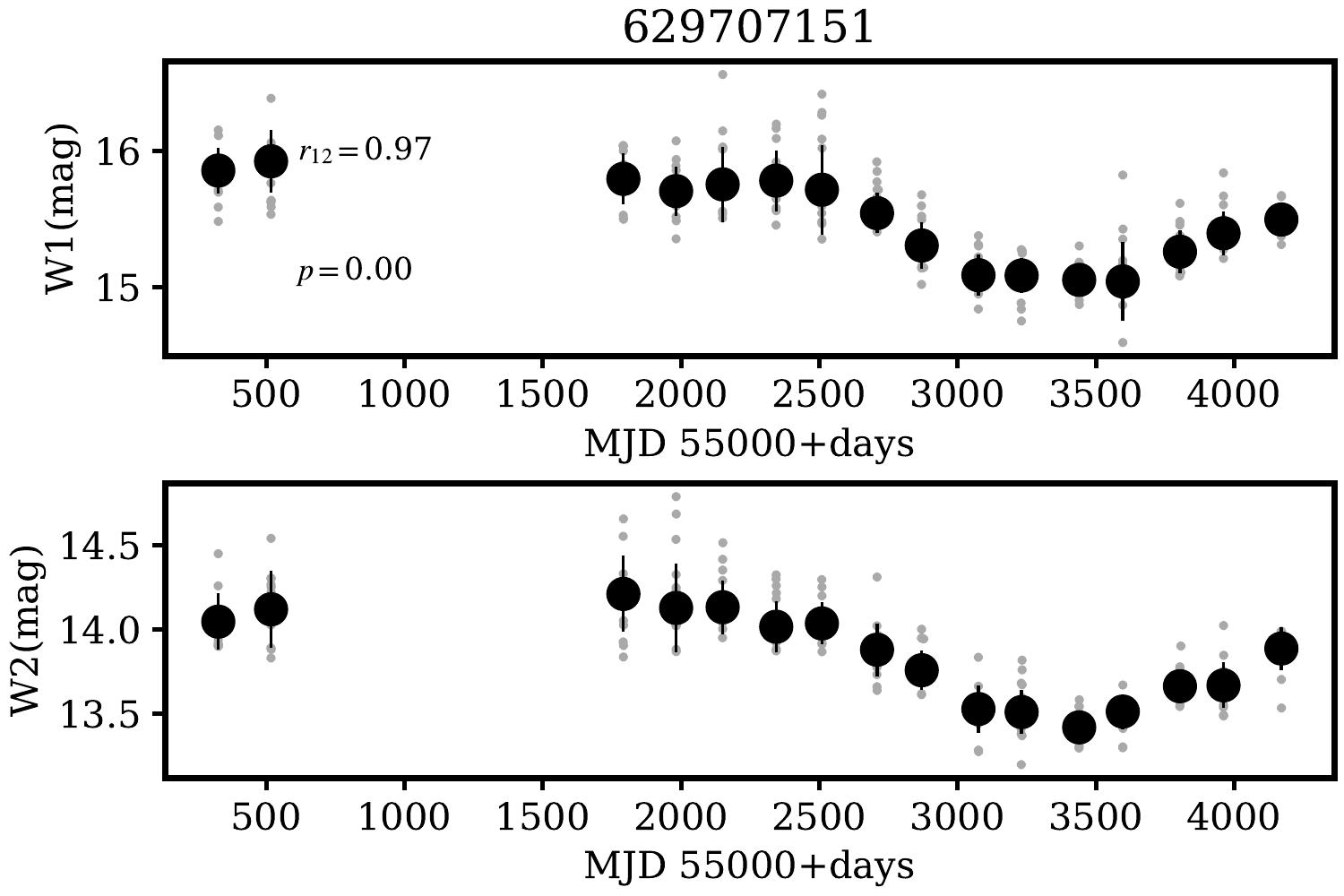}
      \caption{Same as in Fig.\ref{fig:light_curve_1} but for LAMOST obsid 629707151, where the Pearson $r_{12}$ correlation is 0.97.}
         \label{fig:light_curve_2}
  \end{figure}

We have confirmed the existence of AGNs from the CIGALE multi-band photometry fitting result, where we use \citet{2006MNRAS.366..767F} model.  
Among the 138 sources that have multi-wavelength photometry, we have 2 sources where the AGN fraction is over 20\%.
An example of the SED fitting result is shown in Fig.\ref{fig:CIGALE-sed}.  % and Fig.\ref{fig:CIGALE-sed-agn_2}.
%  \begin{figure}[h!]
%   \centering
%   \includegraphics[width=\hsize]{866815044_best_model.pdf}
%       \caption{CIGALE SED fitting with GALEX, SDSS $ugriz$ and WISE $W_1$ to $W_4$ photometry. 
%       The dust temperature with the best fitting result is 200K and the AGN fraction is 21\%.}
%          \label{fig:CIGALE-sed-agn_2}
% \end{figure}
% 
For the MIR color selection, we follow the criteria in \citet{2011ApJ...735..112J} Equation 1 and have obtained 23 AGN candidates under this criteria. %with the follo%and display the result in Fig.\ref{fig:WISE_color_color}.  
% There are 23 sources that fall in this category.
% \begin{equation}
%     \begin{split}
%       \rm  W_1 - W2 &> 2.2 ,\\
%       \rm W_2 - W_3 & > 4.2 ,\\
%       \rm W_1 - W_2 & > 0.1\time (W_2 - W_3)+ 0.38,\\
%       \rm W_1 - W_2 &< 1.7 .
%     \end{split}
% \end{equation}
%  \begin{figure}
%   \centering
%   \includegraphics[width=\hsize]{W1_W2_W3_color_color_plot_wise_140.pdf}
%       \caption{The MIR color-color plot to select AGN. The blue box marks the AGN selection criteria from \citet{2011ApJ...735..112J}.}
%          \label{fig:WISE_color_color}
%   \end{figure} 
Multi-epoch exposures in the WISE catalog can be obtained from AllWISE Multiepoch Photometry Table and NEOWISE-R Single-Exposure Source (L1b) Source Table.
Some filters are applied to select out the reliable detections in the good quality frames, explained in detail in the Explanatory Supplements\footnote{\url{https://wise2.ipac.caltech.edu/docs/release/neowise/expsup/sec2\_3.html} and \url{https://wise2.ipac.caltech.edu/docs/release/allwise/expsup/sec3\_2.html}.  Here we select the sources with \texttt{quality\_score>0}, \texttt{qi\_fact >0}, \texttt{saa\_sep >0}, \texttt{moon\_masked=0}, \texttt{cc\_flags=0}, \texttt{w1rchi2\_ep<2} and \texttt{w2rchi2\_ep<2}.}.   The light curves of the selected sources are displayed in Fig.\ref{fig:light_curve_1} and Fig.\ref{fig:light_curve_2}.  
%Fig.\ref{fig:MIR_variability} demonstrates the distribution of the Pearson $r_{12}$ of $W_1$ and $W_2$ light curves.  
There are 23 sources with the $r_{12}>0.6$ that meet the criteria in \citet{2021arXiv210513400H} as MIR variables and cover the WISE and NEOWISE detections of at least 9 epochs.
% \begin{figure}
% \centering
% \subfloat{\includegraphics[width=\hsize]{W1_W2_light_curve_743708112_3arcsec_filter.pdf}}
% \qquad
% \subfloat{\includegraphics[width=\hsize]{W1_W2_light_curve_629707151_3arcsec_filter.pdf}}
% \label{fig:light_curve}
% \caption{The light curve of one AGN candidate of $W_1$ (upper panel) and $W_2$ (lower panel), where the title marks the obsid from LAMOST.
%       The Pearson $r_{12}$ correlation is 0.93 with p-value of 0.0.
%       The single exposure magnitudes are marked with gray scatter points, where the median value and the standard deviation of the exposures from the same epoch within 60 days are marked with black errorbars.}
% \end{figure}
% \begin{figure}[h!]
%   \centering
%   \includegraphics[width=\hsize]{r12_hist_20211106.pdf}
%       \caption{The distribution of the Pearson $r_{12}$ of the $W_1$ and $W_2$ light curve.}
%          \label{fig:MIR_variability}
%   \end{figure}
   
Furthermore, we have identified a strong X-ray emission source from the strong [OIII]$\lambda5007$ emission line galaxies after cross-matching with 4XMM-Newton DR11\citep{2020A&A...641A.136W,2020A&A...641A.137T}.  This source has the EPIC (indicates parameters combined from those from the available cameras) broadband energy at 0.2 - 12.0 keV of $1.033\times 10^{43}$ erg/s, which is higher than the luminosity of the AGNs in nearby galaxies addressed in \citet{2019A&A...622A..29M}.
The $L_{0.5-4.5keV}=9.27\times 10^{42}$ erg/s higher than the AGN identification criteria for X-ray source in Section 4.4 of \citet{2011ApJS..195...10X} that a source with an intrinsic X-ray luminosity of $L_{0.5-8keV}=3\times 10^{42}$ erg/s is identified as a luminous X-ray AGN.  Located at $z=0.6022$, this source has the EP 9 broad-band energy covering the energy range of 0.5 - 4.5 keV in the observed frame with $\rm L_{0.8-7.2keV}=9.26\times10^{42} erg/s$, which is a lower limit for the $\rm L_{0.5-8keV}$.  The SFR is from the FUV flux is 33.27 $\rm M_{\odot}$/yr and $\rm L_{0.5-8keV}/SFR>41.44$.  Comparing this source with the \citet{2016MNRAS.457.4081B} $\rm L_{0.5-8keV}/SFR$ level of 39.85 with a scatter of 0.25 dex, which is already higher than the \citet{2012MNRAS.419.2095M} relation of 39.59, our source is much higher.  
  
We summarize the sources that have been identified as AGNs with over four of the methods mentioned above in Table \ref{tab:AGN}.  12 galaxies that have been identified with three the methods (8 identified by [MgII]$\lambda2008$ emission lines, MIR color and MIR variability, and 4 galaxies identified with BPT diagram, MIR color, and MIR variability).  For the galaxies that are identified with one or two methods, there are 11 spectra (9 galaxies) that have been identified with the existence of [MgII]$\lambda2800$ line.
We consider all the sources that are identified as AGNs with more than 3 methods or have been identified with the existence of [MgII]$\lambda2800$ line as AGN candidates and exclude these 25 spectra (23 galaxies) from the star formation rate discussions in the following section.

\begin{deluxetable*}{@{\extracolsep{4pt}}cccccc@{}}[!ht]
\tablecaption{AGN candidate identified with over 4 criteria \label{tab:AGN}}
\tablehead{\colhead{LAMOST designation}& \colhead{[MgII]$\lambda$2008}& \colhead{BPT}&\colhead{CIGALE AGN fraction} & \colhead{MIR color criteria} & \colhead{MIR variability} }%\\ \hline
\startdata
% 1237665583788065061 &SF &3 times & 0.02 & in & $N_{ep}=17$, $\sigma_1 >0, \sigma_2 > 0$, $r_{12}=0.21$, \\
J234141.49+140028.1  &1 time &SF & 0.20 & in & $N_{ep}=17$, $\sigma_1 >0, \sigma_2 > 0$, $r_{12}=0.64$, \\
J021459.09-014459.2  &1 time &SF & 0.21 & in & $N_{ep}=17$, $\sigma_1 >0, \sigma_2 > 0$, $r_{12}=0.61$, \\
\enddata
\end{deluxetable*}

\section{Physical properties}
\label{sec:physical_prop}
 
\subsection{Color excess from the Balmer decrement}
\begin{figure}[!ht]
  \centering
  \includegraphics[width=\hsize]{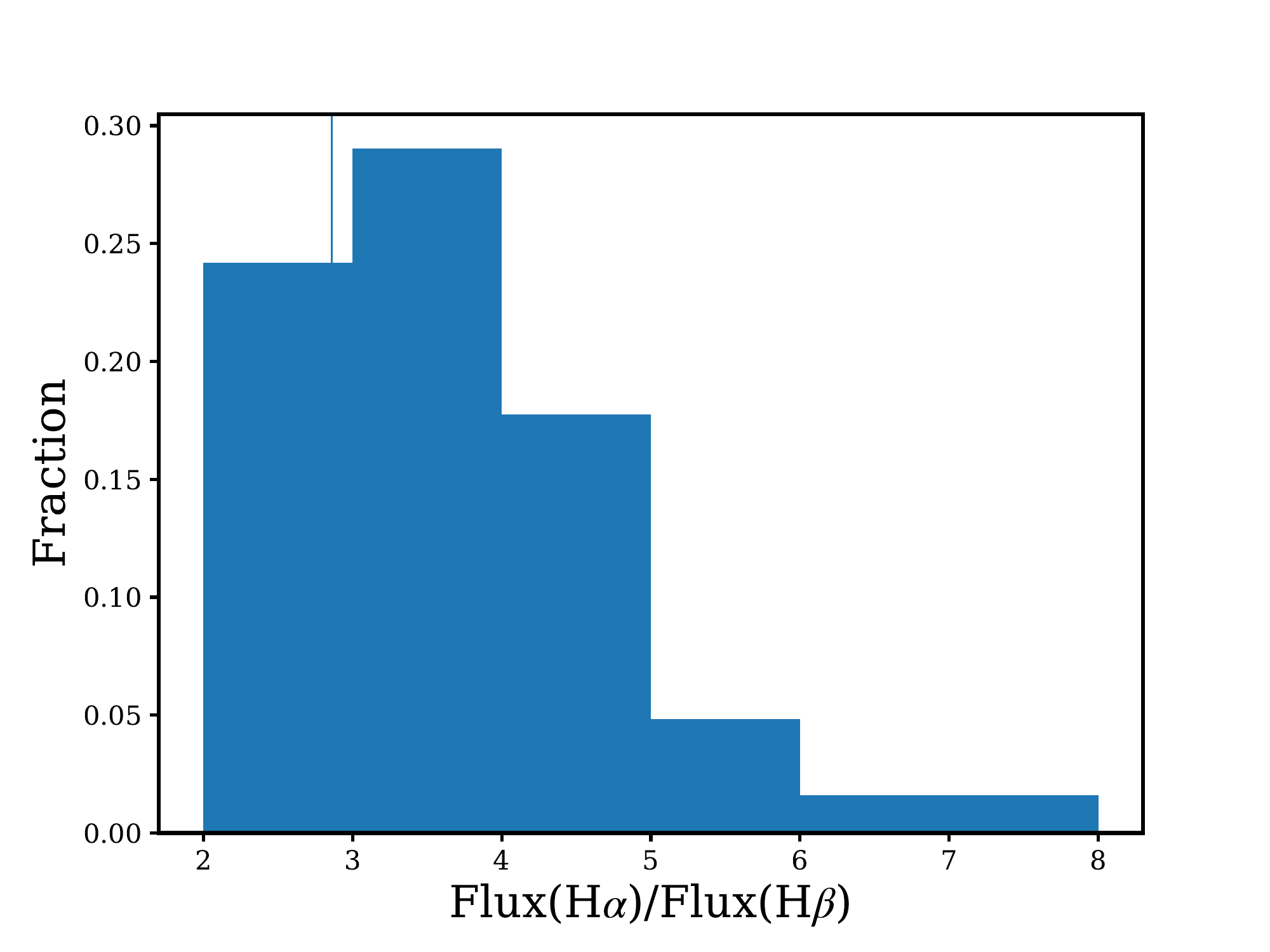}
      \caption{
      The distribution of the flux ratio of $H\alpha/H\beta$.}
         \label{fig:line_ratio}
  \end{figure}

\begin{figure}[!ht]
  \centering
  \includegraphics[width=\hsize]{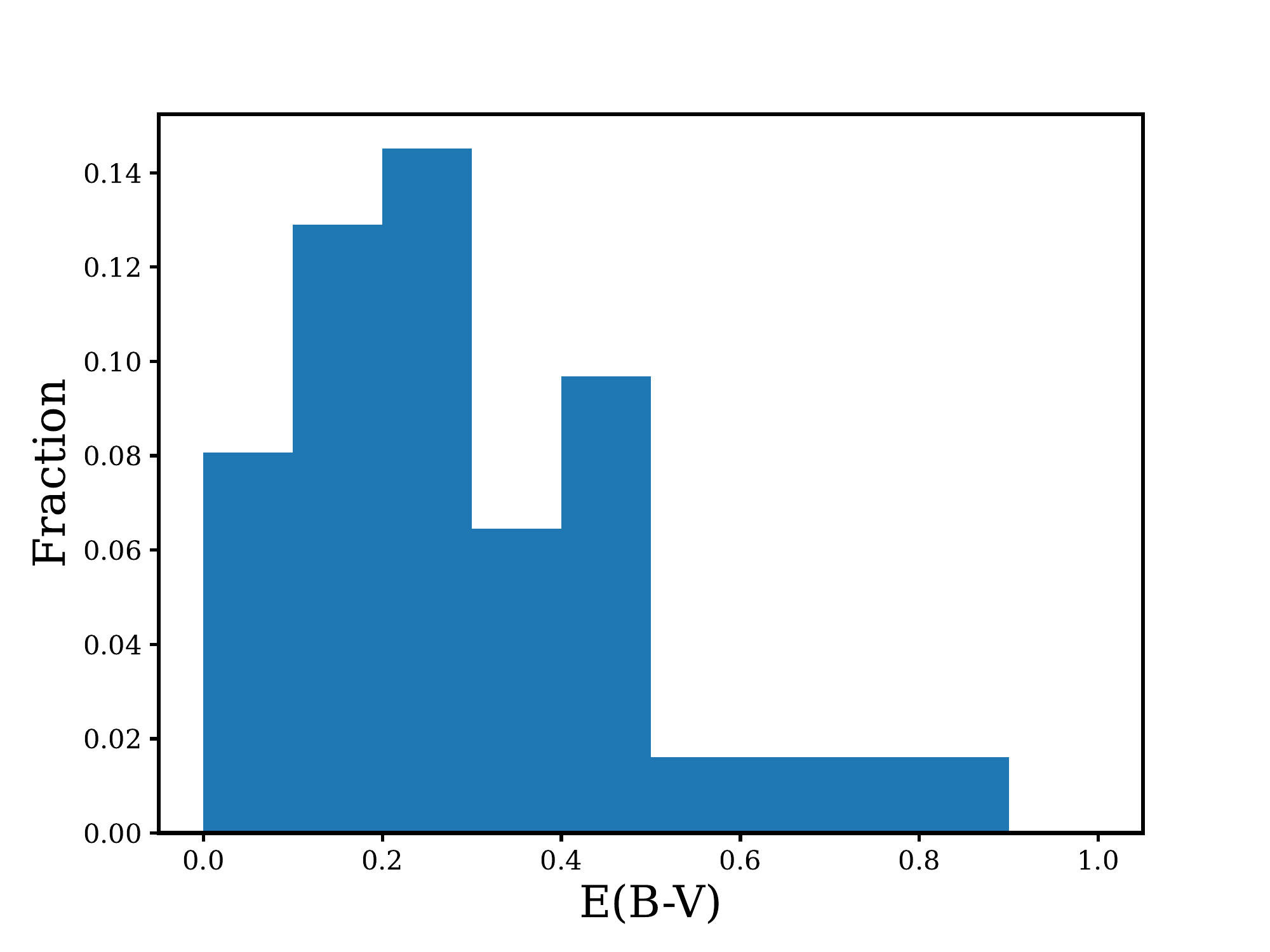}
      \caption{The distribution of the color excess determined from Equation 2.}
         \label{fig:color_excess}
  \end{figure}

% \label{sec:color-excess}
To ensure the quality of the flux ratio measurement, we select the sources at $0.173\le z \le 0.385$ where both the H$\alpha$ and the H$\beta$ emission line locate within the red portion of the spectrograph.
To ensure the validness of the detection, we apply a SNR cut (median value of the flux over the flux error of the H$\beta$ region) over 3 to select the sources.  62 spectra satisfy this criterion, and their flux ratio are displayed in Fig.\ref{fig:line_ratio}.
We measure the color excess of the galaxies from the flux ratio of the H$\alpha$ to H$\beta$ assuming the Case B recombination with the intrinsic line ratio of 2.86.
The color excess from the flux ratio is calculated with:
\begin{eqnarray}
  \rm   E(B-V)_{gas} = \frac{\log_{10}[(f_{H\alpha}/f_{H\beta})/2.86]}{0.4\times [k(H\beta) - k(H\alpha)]},
\end{eqnarray}
where $\rm k(H\alpha)=3.33$ and $\rm k(H\beta)=4.6$ as \citet{2019ApJ...872..145J}.
We assume that the nebular gas in these galaxies emit at $\rm T=10^4K$ and $\rm n_e=10^4 cm^{-3}.$
We show the distribution of the color excess in Fig.\ref{fig:color_excess}.
As is discussed in Section 4.1 in \citet{2009MNRAS.399.1191C}, some of the sources have a flux ratio below 2.86, we contribute this to the uncertainties in flux calibration and low extinction and manually set their color excess to be 0.  

\subsection{Mass and SFR}
The main goal for this subsection is to investigate the mass-metallicity relation and compare the SFR derived from H$\alpha$ emission line and the FUV.  
We use 262 spectra (252 galaxies) with valid Starlight spectral fitting results for deriving the mass-metallicity relation, among which 4 are AGN candidates, 67 spectra (64 galaxies) are Blueberry galaxies, 123 spectra (116 galaxies) are Purple Grape galaxies ($0.05 \le z < 0.112$), and 68 spectra (68 galaxies) are Green Pea galaxies.
We use 24 samples with positive color excess from the previous subsection and FUV detection from GALEX to derive the extinction corrected SFR.

\begin{figure}[!ht]
  \centering
  \includegraphics[width=\hsize]{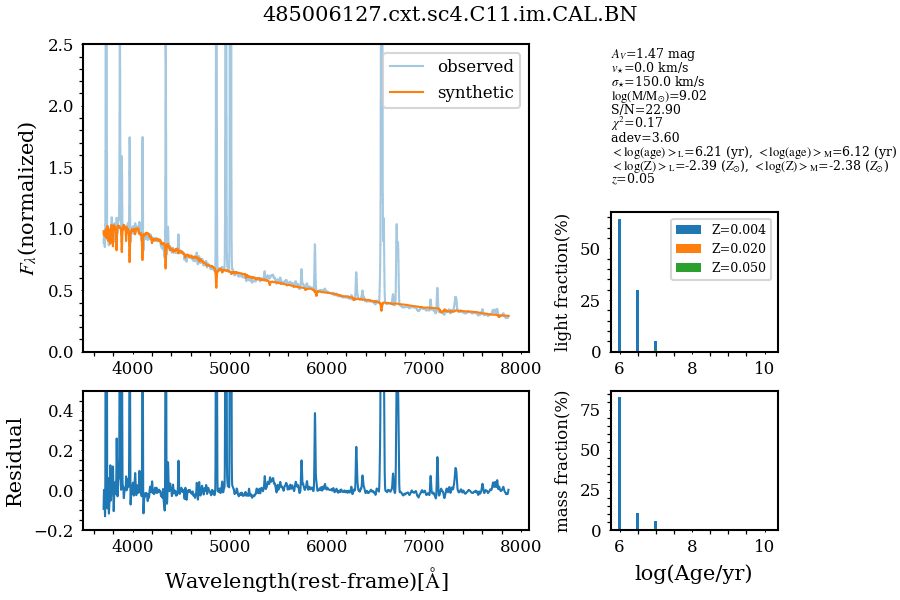}
      \caption{Example Starlight SED fitting result.}
         \label{fig:starlight-sed}
  \end{figure}

To derive the physical properties for these sources \citep{2005MNRAS.358..363C,2006MNRAS.370..721M}, we fit the re-calibrated LAMOST spectra with the emission lines masked.  
The fitting procedure is based on a base set of 45 spectra of three metallicities and 15 ages, with the stellar population model based on the BC03 high resolution spectra, and the \citet{calzetti2000dust} extinction curve.
% We fit the spectra with the emission lines masked.
The initial values of the velocity shift is set to be $v_0=0.0$ km/s, and the velocity dispersion is of $v_d=150.0$ km/s.
We only consider the Starlight fitting of the LAMOST spectra a valid fitting result, where $\rm \frac{\chi^2}{N(\lambda_{eff})}<0.2$, $\rm adev<30$ and SNR ($\rm 6000\AA < \lambda_{rest-frame}< 6500\AA) \ge 2.0$ and have obtained 262 sources.
Fig.\ref{fig:starlight-sed} displays an example of Starlight SED fitting result.
% We use the light-weighted age for following discussions.
%We compare the light-weighted age derived from our Starlight spectral fitting result with \citet{2021arXiv211011610W} 

\begin{figure}[!ht]
  \centering
  \includegraphics[width=\hsize]{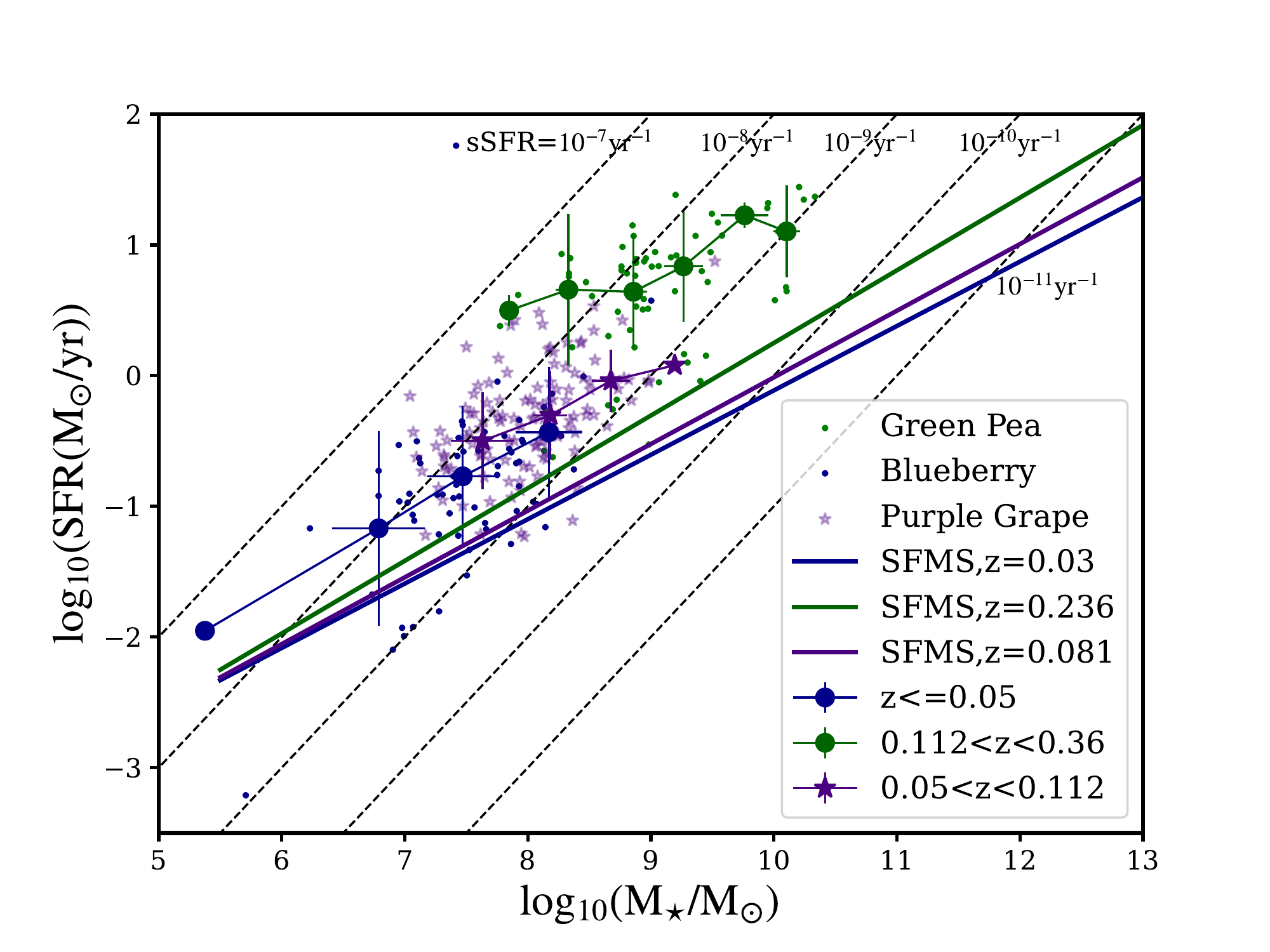}
      \caption{The stellar mass vs. star formation rate (SFR) for our selected strong [OIII]$\lambda5007$ samples.  Blueberry, Green Pea, and Purple Grape galaxies are marked with corresponding colors.  We also mark the Star-Forming Main Sequence (SFMS) relation from \citet{2014ApJS..214...15S} at different redshifts with dark blue ($z=0.03$), dark green ($z=0.236$) solid lines, and indigo lines ($z=0.081$, we only have Purple Grape galaxies with Starlight derived mass in the $0.05 \le z < 0.112$ redshift bin).  The majority of our samples have a higher SFR region compared with the SFMS.}
         \label{fig:mass-sfr}
  \end{figure}

Fig.\ref{fig:mass-sfr} demonstrates the relation of the stellar mass of the galaxies derived from Starlight with SFR (measured from the H$\alpha$ luminosity without extinction correction) \citep{1994ApJ...430..163D,1998ARA&A..36..189K,2003A&A...409...99P,2005AIPC..761..203D,2012ARA&A..50..531K}.
We separate the strong [OIII]$\lambda4363$ emission compact galaxies into Blueberry galaxies, Green Pea galaxies, and Purple Grape galaxies, and compare the mass-SFR relation for these three sets of samples with the SFMS
Within each redshift bin, we separate the sources into several mass bins and calculate the median value and the standard deviation of the mass and metallicity for the sources within this mass and redshift bin.
The main sequence SFR-$M_{\star}$ relation from \cite{2014ApJS..214...15S} is as follows:
% \begin{equation}
% \begin{split}
%     \log \rm SFR (M_{\ast}, t) =& (0.84 \pm 0.02 - 0.026 \pm 0.003 \times t )\log M_{\ast} \\
%     &\quad - (6.51 \pm 0.24 -0.11 \pm 0.03 \times t),
% \end{split}
% \end{equation}
\begin{eqnarray}
    \log \rm SFR (M_{\ast}, t) &=& (0.84 \pm 0.02 - 0.026 \pm 0.003 \times t )\log M_{\ast} \nonumber\\
   & &\quad - (6.51 \pm 0.24 -0.11 \pm 0.03 \times t),
\end{eqnarray}
where $t$ is the age of the universe in Gyr. 
Similar to the discussions in Section 4.4 in \citet{2009MNRAS.399.1191C}, our strong [OIII]$\lambda5007$ samples have much higher sSFR at this redshift range, where typically for the galaxies at $z\sim 0.2$ the sSFR is around $10^{-9} \rm yr^{-1}$ \citep{2004MNRAS.351.1151B,2005ApJ...621L..89B}. 

With the H$\alpha$ emission line measurement result and the FUV flux, we intend to use this sample to check if the SFRs derived from these two indicators agree.
The calculation of the SFR from the recombination emission lines (24 spectra which do not have H$\alpha$ spectral coverage are dropped out) and FUV flux follow the relation defined in  \citet{2011ApJ...741..124H,2011ApJ...737...67M,2012ARA&A..50..531K}, where the corresponding parameters are in Table \ref{tab:opticalsurvey}:
\begin{eqnarray}
\rm \log \dot{M}_{*}(M_\odot year ^{-1} ) = \log L_x - \log C_x.
\end{eqnarray}

\begin{deluxetable}{@{\extracolsep{4pt}}ccc@{}}[!ht]
\tablecaption{Star Formation rate calibrations\label{tab:opticalsurvey}}
\tablehead{\colhead{Band} & \colhead{$L_x$ units} & \colhead{$\log C_x$}}%\\ \hline
\startdata
FUV & ergs $s^{-1}$ ($\nu L_{\nu}$) & 43.35 \\
H$\alpha$  & ergs $s^{-1}$ & 41.27 
\enddata
\end{deluxetable}
% Fig \ref{fig:sfr-sfr} shows the comparison of SFRs derived from these two methods.
% The scatter fit the linear relation well.  
% \begin{figure}[h!]
%   \centering
%   \includegraphics[width=\hsize]{SFR_comparison_20211111_uncorrected.pdf}
%       \caption{SFR (H$\alpha$) vs SFR (FUV).  The black solid line marks the linear relation.  No extinction correction was  applied, because there are only 62 galaxies that have color excess information and these strong [OIII]$\lambda$5007 emission-line galaxies typically are with low-extinction values. }
%          \label{fig:sfr-sfr}
%   \end{figure}
  
As discussed in Section 4.2, we convert from $E(B-V)$ to $A_V$ using $A_V=R_V\times E(B-V)$ to the emission lines for the 62 high H$\beta$ SNR galaxies, under the assumption of \citet{2000ApJ...533..682C} extinction curve and $R_V=3.1$.
Within these 62 samples, 36 samples are with positive color excess.
24 out of the 36 samples have GALEX FUV detections.
Fig.16 illustrates the comparison of SFR without extinction correction (left panel) and after extinction correction (right panel) with these 24 samples. 
\begin{figure*}
\centering
\label{fig:sfr-sfr_corr}
\subfloat{\includegraphics[width=0.45\hsize]{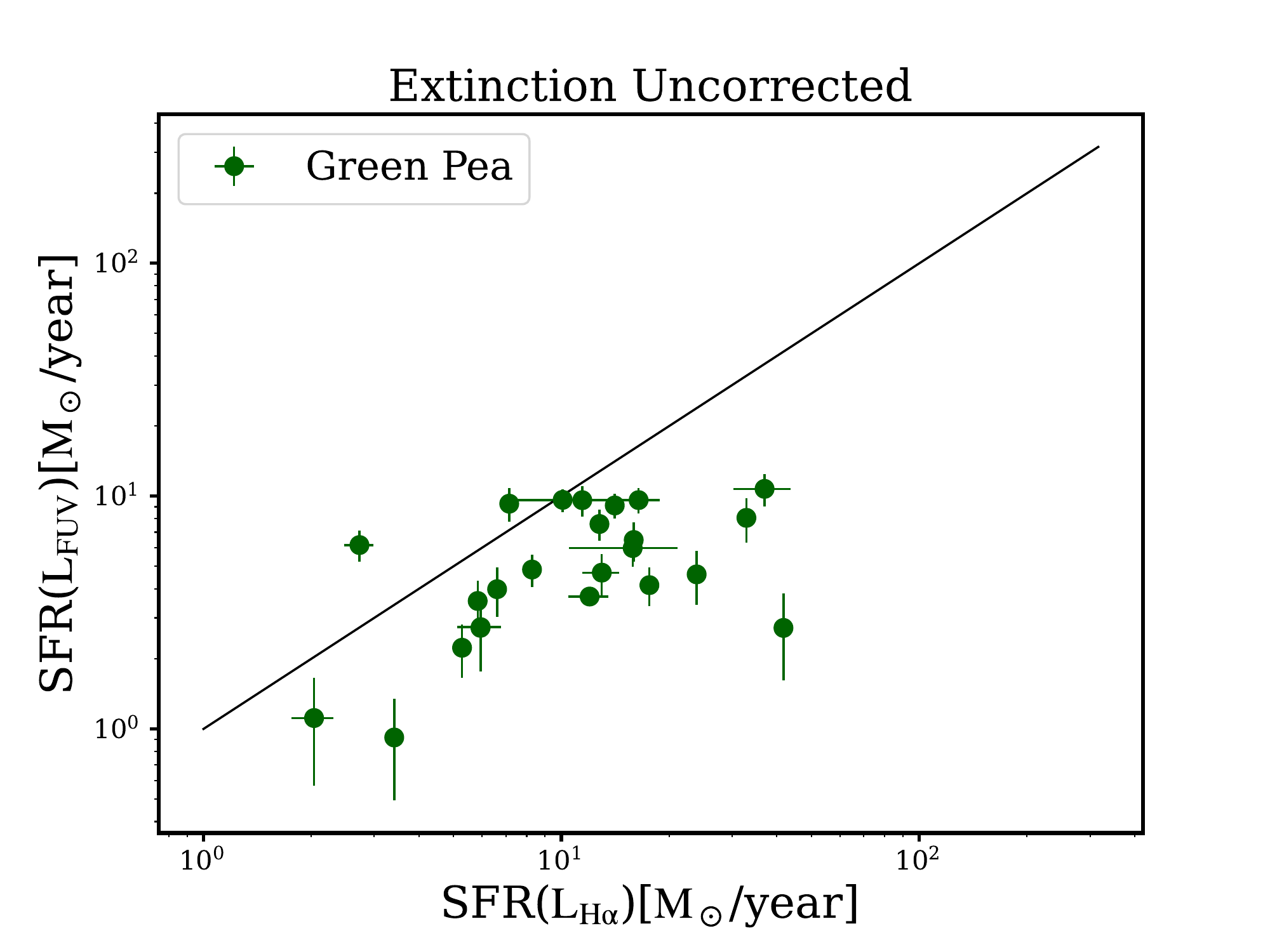}}
\qquad
\subfloat{\includegraphics[width=0.45\hsize]{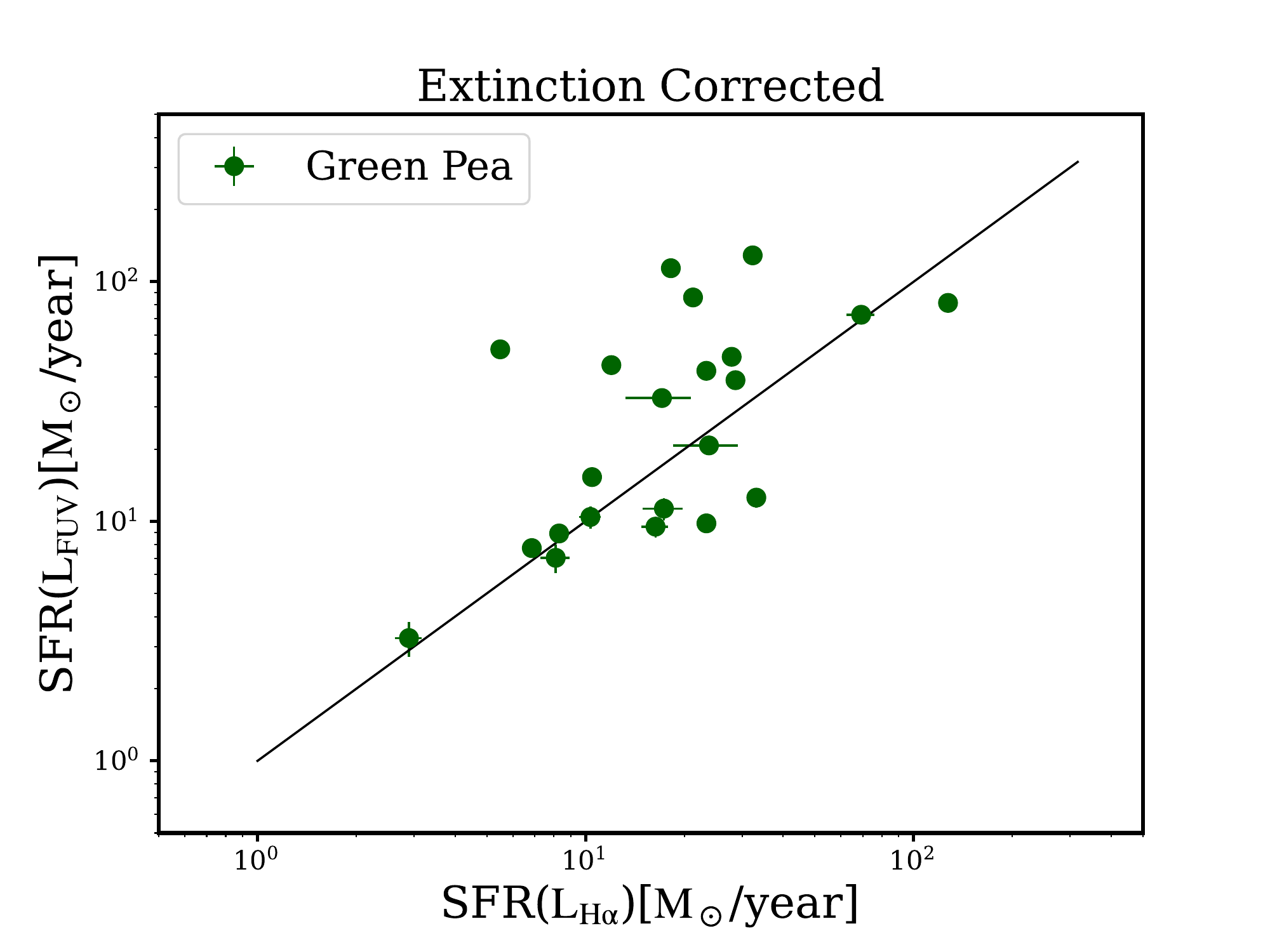}}
\caption{SFR (H$\alpha$) vs SFR (FUV).  
The black solid line marks the linear relation.  
The left panel shows the SFR without attenuation correction and the right panel shows the SFR with attenuation correction.}
\end{figure*}

\subsection{Gas-phase metallicity}
In this subsection, we first show the distribution of metallicity measured from the N2-based method with 1337 spectra, where the [NII]$\lambda6585$ emission line is covered within the spectral range. 
Then we demonstrate the mass-metallicity relation of 252 sources that have Starlight-derived mass and the N2-based metallicity.
Furthermore, we discuss the gas-phase metallicity calculated from the empirical N2-based method and the direct $T_e$-based method with 21 galaxies whose specific flux of the [OIII]$\lambda4363$ emission lines are above $\rm 8.85\times10^{38} erg/s/\AA$, where the detection of [OIII]$\lambda4364$ is rare \citep{2017RAA....17...41G}.

\begin{figure}[!ht]
  \centering
  \includegraphics[width=\hsize]{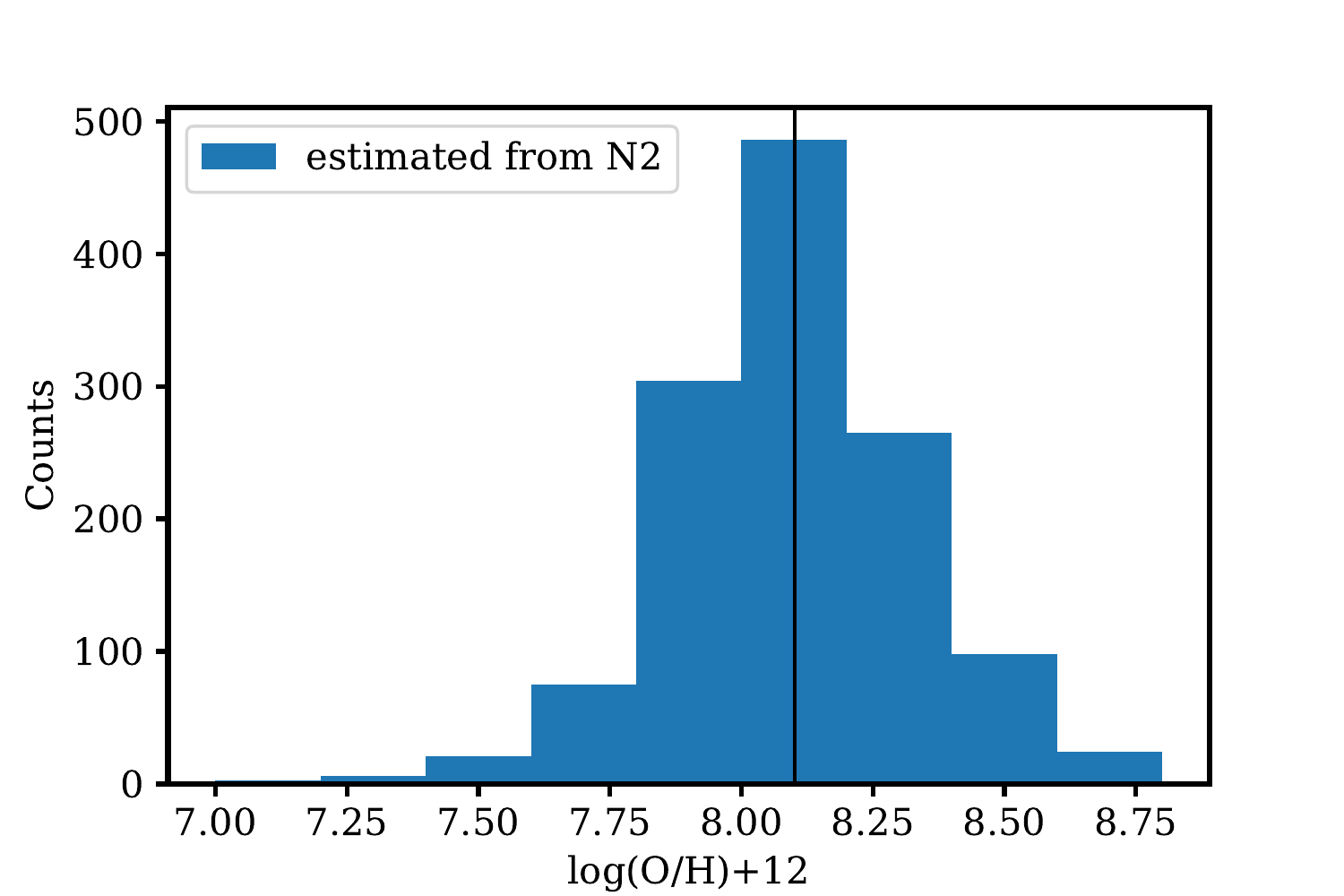}
      \caption{Gas-phase metallicity estimated from the N2-based method.  We mark the median value of the metallicity distribution with the vertical line where $12+\log(O/H)=8.10$, about 0.26 solar metallicity.   }
         \label{fig:N2}
  \end{figure}

We use the N2-based method \citep{1998AJ....116.2805V,2004MNRAS.348L..59P,2013A&A...559A.114M} to empirically calculate the gas-phase metallicity.
Because the two lines used in this method are close, thus less prone to the flux calibration error.
The definition for this index is as follows:
\begin{eqnarray}
    \rm N2 = \rm \log_{10}([NII]\lambda6585/H\alpha),
\end{eqnarray}
and this index can be converted to gas-phase metallicity with the following relation:
\begin{eqnarray}
\rm 12+\log(O/H) = \rm 8.90 + 0.57 \times N2.
\end{eqnarray}
Fig.\ref{fig:N2} displays the distribution of the N2-based method derived gas-phase metallicity.  The gas-phase metallicity is higher than that of the pure Blueberry galaxies \citep{2017ApJ...847...38Y}, similar to that of \citet{2009MNRAS.399.1191C}.
  
\begin{figure}[!ht]
  \centering
  \includegraphics[width=\hsize]{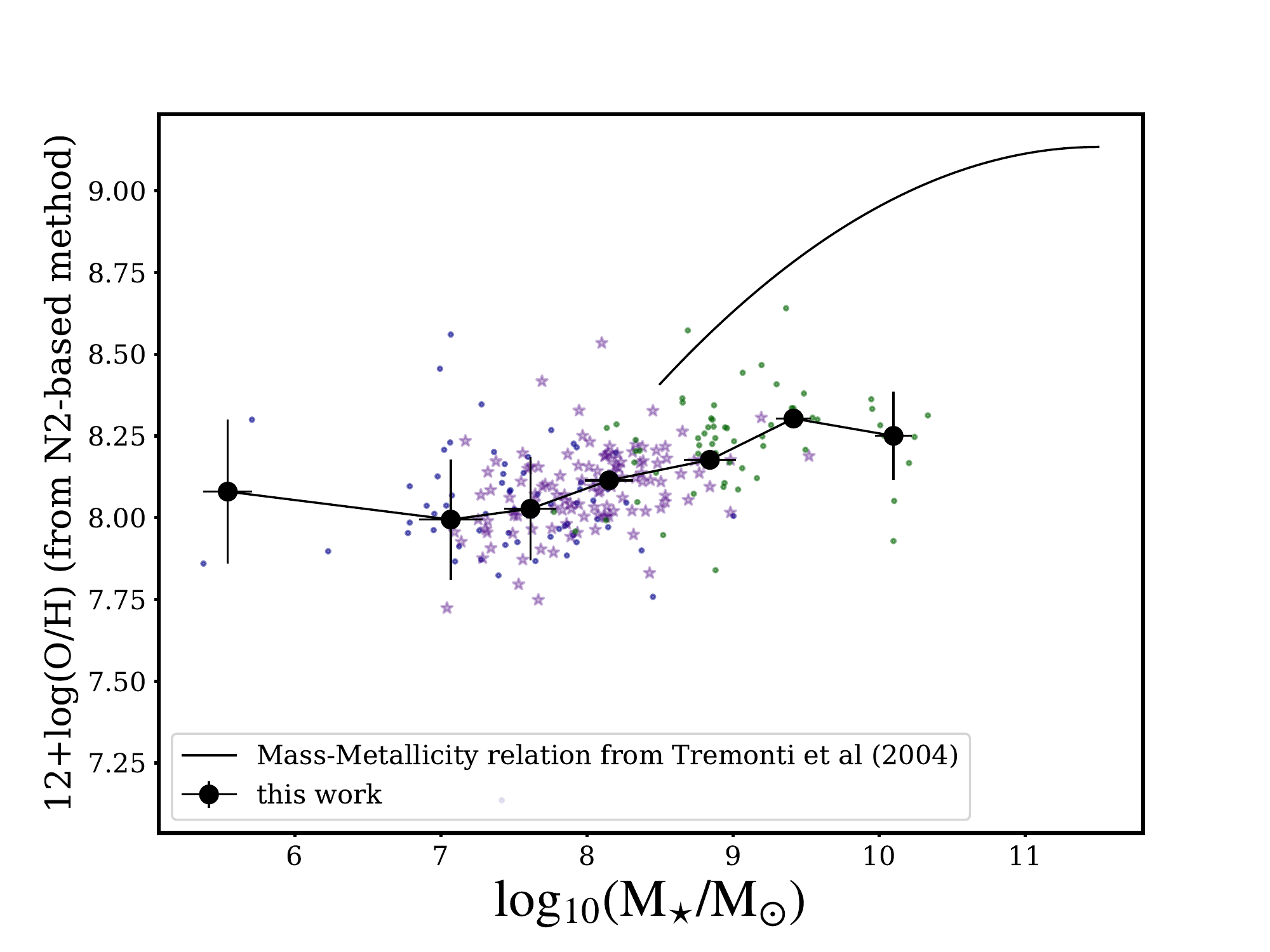}
      \caption{The mass-metallicity relation of 258 spectra with valid Starlight mass measurement.  The blue dots mark the measurements from the Blueberry galaxies, the indigo stars mark the measurements from the Purple Grape galaxies at $0.05 \le z < 0.112$ and the green dots mark the measurements from the Green Pea galaxies. 
      We separate the whole sample into seven mass bins and mark the median value and the standard deviation of the mass and metallicity with black error bars.
      The black solid curve on the top right is from \citet{2004ApJ...613..898T}.}
         \label{fig:mass-metal}
  \end{figure}

We show the distribution of the mass-metallicity relation in Fig. \ref{fig:mass-metal}.  %%% add comments here.
% As in Figure 15, we separate the strong [OIII]$\lambda4363$ emission compact galaxies into Blueberry galaxies, Green Pea galaxies and Purple Grape galaxies, and compare the mass-metallicity relation for these three sets of samples.
We separate 248 galaxies (258 spectra), which consist of 64 Blueberry galaxies (67 spectra), 68 Green Pea galaxies (68 spectra), and 116 Purple Grape galaxies (123 spectra) into seven mass bins and calculate the median value and the standard deviation of the mass and metallicity for the sources within this mass bin.
% We separate the groups of galaxies into three bins

We demonstrate that the metallicity increases with the mass by comparing the points, with the smallest mass bin and the largest mass bin excluded where are fewer data points in these two mass bins.
All of these strong [OIII]$\lambda4363$ emission line compact galaxies are below the \citet{2004ApJ...613..898T} mass-metallicity relation.
% Comparing the three curves, we demonstrate that the mass and metallicity are increasing with the redshift.

Besides the empirical method to calculate the gas-phase metallicity, there is also the direct $T_e$-based method.  However, the auroral [OIII]$\lambda4363$ line is not easy to detect.
In this work, 21 galaxies whose specific flux of the [OIII]$\lambda4363$ emission lines are above $\rm 8.85\times10^{38} erg/s/\AA$ are used to test the direct $T_e$-based method to calculate the electron temperature and the metallicity referring to \citet{2006A&A...448..955I} Section 3.1 and \citet{2019ApJ...872..145J}.
The purpose of this discussion is to show that we have detected the auroral [OIII]$\lambda4363$ emission line, which increases the samples of detections in LAMOST \citep{2017RAA....17...41G}, and to demonstrate the current N2-based empirical metallicity relation and direct $T_e$-based metallicity relation in \citet{2019ApJ...872..145J} are in agreement within 1$-\sigma$ level.

This approach assumes two electron temperatures for $\rm O^+$ and $\rm O^{++}$ in a two-zone photo-ionization model.
We calculate the $\rm O^{++}$ electron temperature with the following equation, (as in \citet{2006A&A...448..955I} Equations (1) and (2)):
\begin{eqnarray}
    t = \frac{1.432}{\log[(\lambda4959 + \lambda 5007)/\lambda4363]-\log C_T},
\end{eqnarray}
where $\rm t=10^{-4}T_e(OIII)$, and 
\begin{eqnarray}
C_T = (8.44-1.09t+0.5t^2-0.08t^3)\frac{1+0.0004x}{1+0.044x},
\end{eqnarray}
where $x=10^{-4}N_et^{-0.5}$.
We estimate electron temperature of [OII] for the low-metallicity situation in \citet{2006A&A...448..955I} Section 3.1 Equation (14) as follows:
\begin{eqnarray}
  \rm T_e(OII) = -0.577 + T_e(OIII)\times (3.065-0.498T_e(OIII)).
\end{eqnarray}
This is also the method that \citet{2019ApJ...872..145J} uses for calculation, and they state that their measurement of the oxygen abundance depends little on the relation between $T_2$ and $T_3$.
Similarly, we use the term$\rm T_2 = 10^{-4}T_e([OII])$ and $\rm T_3 = 10^{-4}T_e([OIII])$ for clarity.
Our spectra do not cover the [SII]$\lambda$6717 or [SII]$\lambda$6731 emission lines.  The derived $\rm N_e$ is always smaller than $\rm 10^3 cm^{-3}$, and as in \citet{2019ApJ...872..145J} $\rm N_e=10, 100 $ or $\rm 10^3 cm^{-3}$ the results do not vary much.
We use $x=0$ for our calculations.

\begin{figure}[!ht]
  \centering
  \includegraphics[width=\hsize]{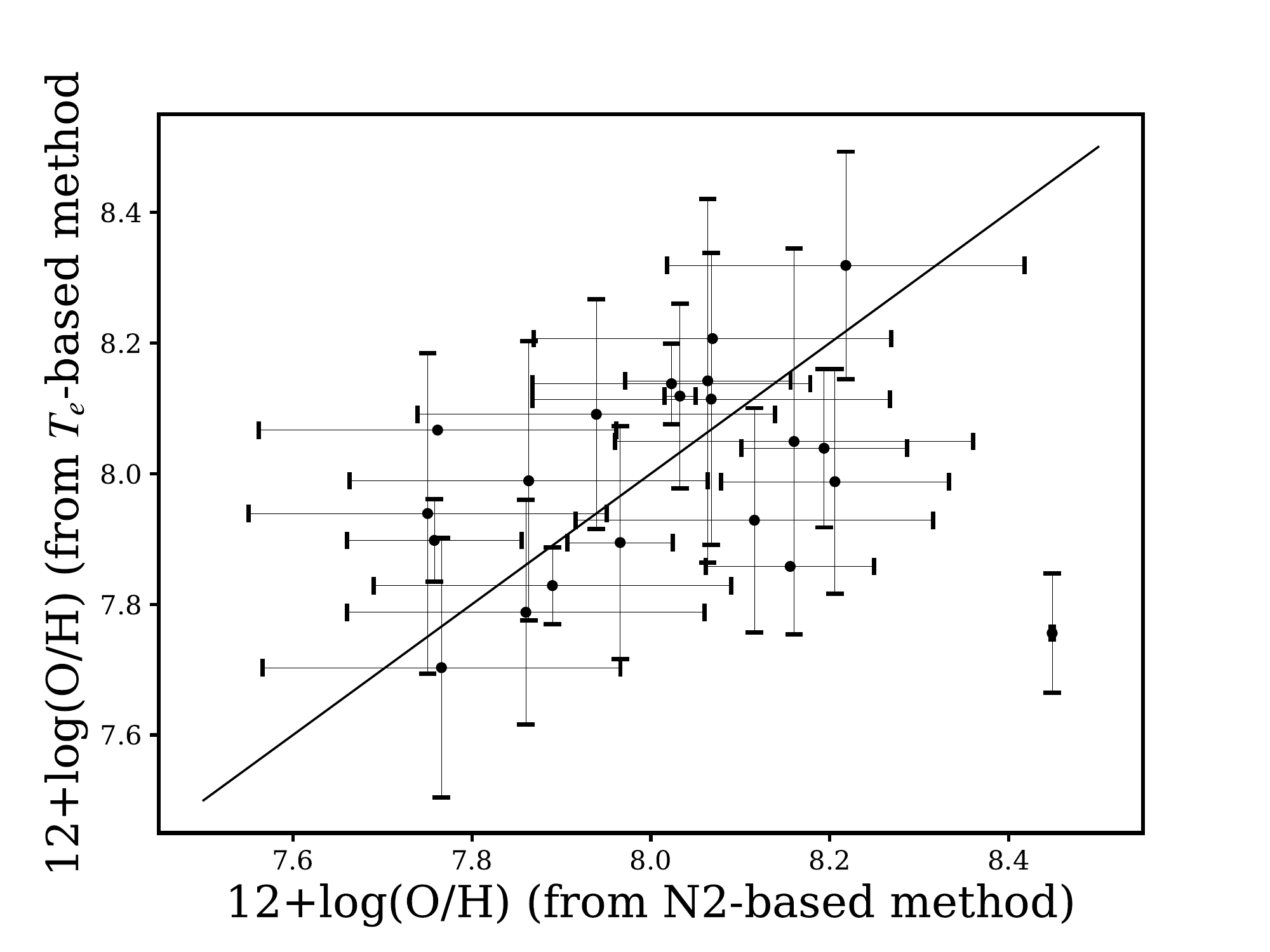}
      \caption{The relation of the derived gas-phase metallicity from the N2-based method (x-axis) and the direct $T_e$-based method (y-axis) is based on our 21 [OIII]$\lambda$4363 samples.  Adopting $\rm 12+\log(O/H)=8.69$ as in \citet{2001ApJ...556L..63A}, these 21 samples are all with sub-solar metallicities.}
         \label{fig:gas-phase-metal}
  \end{figure}

% We use [OII]$\lambda$3727 and [OIII]$\lambda\lambda$4959,5007 for the oxygen abundance calculation.
With the Equations (3) and (5) in \citet{2006A&A...448..955I}, we calculate the ionic abundances as follows:
% \begin{equation}
% \begin{split}
%   \rm  12+\log \frac{O^+}{H^+} =& \rm \log \frac{\lambda 3727}{H\beta} + 5.961 + \frac{1.676}{T_2} - 0.40 \log T_2\\
%     &- 0.034 T_2 + \log(1+1.35x),
% \end{split}
% \end{equation}
\begin{eqnarray}
    \rm  12+\log \frac{O^+}{H^+} &=& \rm \log \frac{\lambda 3727}{H\beta} + 5.961 + \frac{1.676}{T_2} - 0.40 \log T_2 \nonumber\\
   & &- 0.034 T_2 + \log(1+1.35x),
\end{eqnarray}
and 
% \begin{equation}
% \begin{split}
%   \rm  12+\log \frac{O^{2+}}{H^+} =& \rm \log \frac{\lambda 4959 + \lambda5007}{H\beta} + 6.200 + \frac{1.251}{T_3} \\
%     &- 0.55\log T_3 - 0.014 T_3,
% \end{split}
% \end{equation}
\begin{eqnarray}
  \rm  12+\log \frac{O^{2+}}{H^+} &=& \rm \log \frac{\lambda 4959 + \lambda5007}{H\beta} + 6.200 + \frac{1.251}{T_3} \nonumber\\
    & &- 0.55\log T_3 - 0.014 T_3,
\end{eqnarray}
Because the majority of the ions of oxygen are $\rm O^{+}$ and $\rm O^{2+}$, we use $\rm O/H = O^+/H + O^{2+}/H$ to determine oxygen abundance.
Fig.\ref{fig:gas-phase-metal} demonstrates the comparison of the gas-phase metallicity estimated from the N2-based method and the direct $T_e$-based method.
We calculate the error bars in Fig 19 by the Monte Carlo method.  
For the error bars in the N2-based metallicity, we generate 100 realizations for each source within the wavelength range of $6500-6640\AA$ with the flux error from the rescaled LAMOST spectra, and measure the [NII]$\lambda6585$ line and the H$\alpha$ line flux for each realization.  We calculate the N2-based metallicity for each realization and use the standard deviation from 100 realizations as the error bar for that source.
For the error bars in the direct $T_e$-based method, similarly, we generate 100 realizations for each source but in a tighter wavelength range $4320-4383\AA$ and measure the [OIII]$\lambda4363$ line only for each realization.
We note that 15 out of 21 sources locate on the black line within 1$-\sigma$ region.  
This verifies that the current N2-based empirical metallicity relation and the direct $T_e$-based metallicity relation in \citet{2019ApJ...872..145J} are in agreement within 1$-\sigma$ level.

\subsection{Environment}
\begin{figure}[!ht]
  \centering
  \includegraphics[width=\hsize]{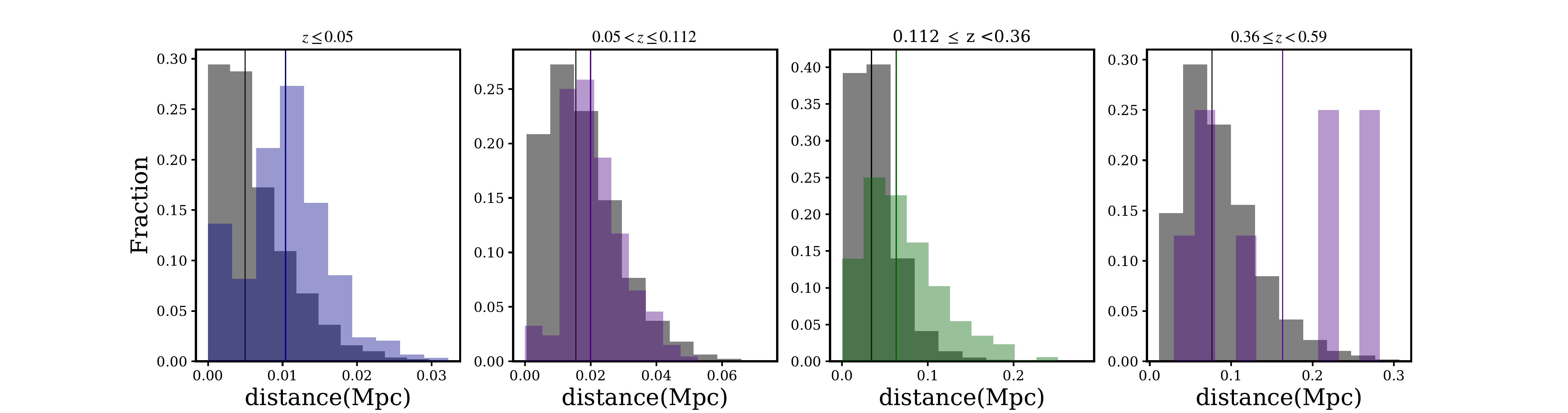}
      \caption{The histogram of the distance to the nearest neighbor.  
The colored histogram displays the strong [OIII]$\lambda5007$ samples while the gray histogram shows the parent comparison star forming galaxies sample selected from SDSS. 
We mark the median values (strong [OIII]$\lambda5007$ samples, parent SFGs) of the two distributions with vertical lines of corresponding colors.}
         \label{fig:distance}
  \end{figure}

We identify a parent sample of comparison star forming galaxies from the ``emissionLinesPort" catalog \citep{2013MNRAS.431.1383T} whose spectroscopic classification is `Galaxy', the BPT classification is `star-forming' and cover the same redshift range as our Blueberry galaxies ($z\leq 0.05$), Purple Grape galaxies ($0.05 <z \leq 0.112$ and $0.36 \leq z < 0.59$), Green Peas ($0.112\leq z<0.36$).
We ensure the parent comparison sample covers the same mag-z space as our selected samples in the $r-$band as displayed in the following plot.
The Neighbors table from SDSS16 provides the angular separation to the nearby objects within 0.5 arc minutes.  We only consider the primary detections as the valid neighbors.  We calculate the projected linear separation between our source and all the valid neighbors by multiplying the angular separation from the Neighbors table with the luminosity distance at the redshift of our strong [OIII]$\lambda5007$ emission line compact dwarf galaxies.  We consider the smallest linear separation as the distance to the nearest neighbor.
We display the distance for our samples to the nearest neighbor in different redshift bins in Fig.\ref{fig:distance}.
% We also show the cumulative count distribution of the neighbor counts for the strong [OIII]$\lambda5007$ samples as well as the parent comparison star forming galaxies sample in Fig.\ref{fig:distance_cumsum}.
%       \begin{figure}[h!]
%   \centering
%   \includegraphics[width=\hsize]{distance_cumcount_20211019.pdf}
%       \caption{The cumulative distribution of neighbour counts within 10 kpc for the strong strong [OIII]$\lambda5007$ samples (solid black line) as well as the parent comparison star forming galaxies sample (dashed gray line).}
%          \label{fig:distance_cumsum}
%   \end{figure}

We also perform the Komogorov-Smirnov test for these four sets of comparisons, and the p-value between the distance to the nearest neighbor of the parent sample and our selected sample in four redshift bins are all smaller than 0.05. 
Therefore, it is apparent that the strong [OIII]$\lambda5007$ samples are further away from the nearest neighbor compared with typical star forming galaxies.
% However, we perform a Komogorov-Smirnov test and the p-value between cumulative number counts of neighbors within 10 kpc of strong [OIII]$\lambda5007$ and the parent SFGs are 0.787, and the p-value of the distance to the nearest neighbors of [OIII]$\lambda5007$ galaxies and the parent SFGs is 0.168.  
% Thus we cannot reject the null hypothesis that these two groups are sampled from populations with different distributions.

% \subsection{Analogs to high$-z$ Lyman-$\alpha$ break galaxies}
% Similar to the ultraluminous compact galaxies, our samples are compact in size, bright in UV luminosity.

\section{Result}
\label{sec:result}
 We have selected 1547 unique strong [OIII]$\lambda$5007 emission line compact galaxies  with 1694 spectra\footnote{\url{https://doi.org/10.12149/101085}} from LAMOST DR9.  1342 galaxies are observed spectroscopically for the first time. Our samples enlarge the current samples of luminous compact galaxies, Green Pea galaxies, and Blueberry galaxies.
The following LAMOST extragalactic survey will keep providing large amount of spectra for these targets and provide duplicate detections for the same source helpful for observing the line variability.

We have confirmed 23 AGNs candidates (25 spectra) from this strong [OIII]$\lambda$5007 emission line compact galaxies by LAMOST optical spectra, BPT diagram, CIGALE SED fitting, MIR color, and MIR variability.

%  Multi-wavelength CIGALE SED fitting result confirm the existence of the host dust in these strong [OIII]$\lambda5007$ emission-line galaxies, which is strongly correlated with $W_2-W_3$ color.  
 
Our samples show that strong [OIII]$\lambda5007$ emission-line galaxies have higher SFR compared with the main-sequence star formation rate.  The SFR from the FUV luminosity with the H$\alpha$ luminosity agrees with the coefficients after extinction correction from \citet{2012ARA&A..50..531K}. 
From the mass-SFR plot, we show that with the increase of redshift, the SFR is increasing.
For the sources within the same redshift bin, the SFR increases with mass with a similar slope as the SFMS.
 
For the gas-phase metallicity for these samples, these samples have a median metallicity of 12+log(O/H) of 8.10.  In the mass-metallicity plot, all the sources are below the \citet{2004ApJ...613..898T} mass-metallicity relation.  The metallicity increases with mass. 
21 galaxies are found with [OIII]$\lambda$4363 emission lines.  The direct $T_e$-based metallicity measurement result is in agreement with the N2-based metallicity result.
 
Lastly, Strong [OIII]$\lambda5007$ emission-line galaxies are in a less dense environment.

% % \newpage

% \begin{acknowledgements}
% \acknowledgements
\section*{Acknowledgements}
This work is supported by National Science Foundation of China (No. U1931209) and National Key R\&D Program of China(No. 2019YFA0405502).  Guoshoujing Telescope (the Large Sky Area Multi-Object Fiber Spectroscopic Telescope, LAMOST) is a National Major Scientific Project built by the Chinese Academy of Sciences.  
Funding for the project has been provided by the National Development and Reform Commission.  
LAMOST is operated and managed by the National Astronomical Observatories, Chinese Academy of Sciences.
Siqi Liu thanks the useful discussion with Sophia Y. Dai, Qing Liu, Er-Lin Qiao and He-Yang Liu.
This research uses Astropy,\footnote{\url{http://www.astropy.org}} \citep{2013A&A...558A..33A,2018AJ....156..123A}, TOPCAT \citep{2005ASPC..347...29T}, Scipy \citep{2020NatMe..17..261V}, Numpy \citep{2020Natur.585..357H} and Matplotlib \citep{Hunter:2007}
% \end{acknowledgements}
% -------------------------------------------------------------------
\newpage
\bibliographystyle{aasjournal}
\bibliography{GP_new.bib}
% \newpage
% \appendix

\end{document}